\documentclass[aps,prd,preprint,superscriptaddress,amsmath,amssymb,showpacs]{revtex4-1}
\usepackage{dcolumn}
\usepackage{graphicx}
\usepackage{float}
\usepackage{physics}
\usepackage[colorlinks=true,allcolors=blue]{hyperref}

\begin{document}

\title{Shear viscosity coefficient of magnetized QCD medium near chiral phase transition}

\author{Xueqiang Zhu}
\affiliation{College of Science, China Three Gorges University, Yichang 443002, China}

\author{Sheng-Qin Feng}
\email{Corresponding author: fengsq@ctgu.edu.cn}
\affiliation{College of Science, China Three Gorges University, Yichang 443002, China}
\affiliation{Key Laboratory of Quark and Lepton Physics (MOE) and Institute of Particle Physics,\\
Central China Normal University, Wuhan 430079, China}
\affiliation{Center for Astronomy and Space Sciences, China Three Gorges University, Yichang 443002, China}

\date{\today}

\begin{abstract}
Abstract: We study the properties of the shear viscosity coefficient of quark matter at finite temperature and chemical potential near chiral phase transition in a strong background magnetic field. A strong magnetic field induces anisotropic features, phase-space Landau-level quantization, and if the magnetic field is sufficiently strong, interferes with prominent QCD phenomena such as dynamical quark mass generation, likely affecting the quark matter transport characteristics. The modified Nambu-Jona-Lasinio (NJL) model with inverse magnetic catalysis effect by fitting the Lattice QCD (LQCD) results is used to calculate the changes of quasiparticle related thermodynamic quantities, and the shear viscosity of the system medium, which is analyzed under the relaxation time approximation. We quantify the influence of the order of chiral phase transition and the critical endpoint on dissipative phenomena in such a magnetized medium. When the magnetic field exists, the shear viscosity coefficient of the dissipative fluid system can be decomposed into five different components. In strong field limit, we make a detailed study of the dependencies of $\eta_{2}$ and $\eta_{4}$ on temperature and magnetic field for the first order phase transition and critical endpoint transition. respectively. It is found that $\eta_{2}$ and $\eta_{4}$ both decrease with magnetic field and increase with temperature, and the discontinuities of $\eta_{2}$ and $\eta_{4}$ occur at the first order phase transition point.
\end{abstract}


\maketitle

\section{Introduction}\label{sec:01_intro}
It is well known that strong magnetic fields can be produced in relativistic heavy-ion collisions \cite{Bzdak:2011yy,Deng:2012pc,Kharzeev:2007jp,Mo:2013qya,Zhong:2014cda,Feng:2016srp}. The magnetic field strengths are comparable to or even larger than the strong-interaction scale $\Lambda_{\textrm{QCD}}\approx0.25~\mathrm {GeV}$. The magnitudes of magnetic field generated for Pb-Pb collisions at the Large Hadron Collider (LHC) are computed as large as $eB\sim15m_{\pi}^{2}\gg\Lambda_{\textrm{QCD}}^{2}$ \cite{Skokov:2009qp}. A strong magnetic field acting on strong QCD leads to some new phenomena, such as chiral anomaly  \cite{Tuchin:2013ie,Kharzeev:2015znc,Guo:2019joy,Zhao:2019hta} in QGP environment. As for the role of magnetic field in Ref. \cite{Kharzeev:2007jp,Tuchin:2013ie}, those who hold opposite views believe that the fields are produced early in the collision and decay fast as the system expands. On the other hand, the early generated fields can induce electric currents in the expanding system which in turn make magnetic fields last longer while the system exists  \cite{Chen:2019qoe,Tuchin:2015oka,McLerran:2013hla,She:2017icp}.

One of the major progresses in the study of heavy-ion collision (HIC) experiments like the Relativistic Heavy Ion Collider (RHIC) and Large Hadron Collider (LHC) is that the generated medium manifested as a nearly perfect fluid \cite{Schafer:2009dj}, with the smallest shear viscosity to entropy density ratio ($\eta/s$) ever discovered in nature. A considerable amount of research has already been performed in studying the influence of the magnetic field on the QCD phase diagram \cite{Andersen:2014xxa}. The modification of the QCD phase diagram in the presence of magnetic field is directly related to the corresponding change in the quark condensate, and its enhancement with magnetic field is known as magnetic catalysis, which is a quite expected feature in vacuum as well as at finite temperature \cite{Klevansky:1989vi,Gatto:2010pt,Miransky:2015ava,Boomsma:2009yk,Gatto:2012sp,Gatto:2012sp,Chatterjee:2014csa}.

Recent lattice QCD results \cite{CMS:2012sap,Bali:2012zg,Bruckmann:2013oba,Endrodi:2013cs} have authenticated that strong magnetic fields have striking effects on the QCD phase diagram, especially in the region approach to the pseudocritical temperature $T_{\textrm{PC}} = 0.17~\mathrm{GeV}$, the region related with the hadron-to-quark transition. The magnetic field effects are most striking on the $u$ and $d$ quark condensates, namely the condensates increase with the magnetic field for low temperatures which refers to magnetic catalysis (MC), and decrease for temperatures close to $T_{\textrm{PC}}$  which refers to inverse magnetic catalysis (IMC) \cite{Andersen:2014xxa,Gatto:2012sp,Miransky:2015ava}.  The quark condensates have a direct impact on the effective masses of the $u$ and $d$ quarks, which in turn play an important role to the transport feature. The modifications pertaining to the QCD phase diagram may also have some impact in the transport properties of the medium produced in HICs.

The shear $\eta$ and bulk $\zeta$ viscosities are parameters that quantify dissipative processes in the hydrodynamic evolution of QCD fluid in the background of magnetic field. Furthermore, it has been indicated that $\zeta$ and $\eta$ are sensitive to phase transitions in a magnetized medium \cite{Sasaki:2008um,Soloveva:2021quj,Mykhaylova:2020pfk}. In QCD of particular interest are properties of transport coefficients near the critical line where deconfinement and chiral phase transition sets in. This is because any change of the bulk and shear viscosities near $T_{c}$ modifies the hydrodynamic evolution of the QCD medium and influences phenomenological observables that characterize its expansion dynamics. Clearly, in the evolution of QCD medium its transport properties are changing when approaching the phase boundary. Thus, it is of importance to quantify the change of shear viscosities with thermal parameters near the phase transition expected in the magnetized QCD medium.

In the present work we will study the shear viscosity of magnetized quark matter near the quark chiral phase transition point by using a dynamical quasiparticle Nambu-Jona-Lasinio (NJL) model \cite{Hatsuda:1994pi,Buballa:2003qv}. Among the earlier study of shear viscosity for magnetized matter  \cite{Li:2017tgi,Nam:2013fpa,Alford:2014doa,Tuchin:2011jw,Hattori:2017qih,Huang:2009ue,Huang:2011dc,Mohanty:2018eja}, one finds that Refs. \cite{Li:2017tgi,Nam:2013fpa,Alford:2014doa} have not studied its component decomposition, which is explored in Refs. \cite{Tuchin:2011jw,Hattori:2017qih,Huang:2009ue,Huang:2011dc,Mohanty:2018eja}. Much of our studying of the low-energy regime of \textrm{QCD}, and of the \textrm{QCD} phase diagram in particular, is built on insights gained with quasiparticle models. The effective quark masses are determined by the in-medium quark condensate. As mentioned, strong magnetic fields change the condensate and in the present paper take this effect into account in the computation of the shear viscosity coefficient. Our proposal is that basically the same effect can be obtained by assuming that the \textrm{NJL} quark-quark coupling decreases with $B$ and $T$ , as does the \textrm{QCD} coupling. As a result, the NJL model results turn out to be in qualitative agreement with the lattice results~\cite{Bali:2011qj,Bali:2012zg}; the condensate grows in accordance with MC and the pseudocritical temperature for the chiral transition decreases with $eB$. And then, we use kinetic model of Boltzmann equation with the relaxation-time approximation to study the characteristics of viscosity coefficient components of the first order phase transition and critical endpoint (\textrm{CEP})  transition regions in the strong magnetic field background.

The paper is organized as follows. In Sec.II we introduce the NJL model quasiparticle description of magnetized quark matter, with particular emphasis on the magnetic catalysis and inverse magnetic catalysis of the chiral condensate by fitting LQCD results. After considering the influence of inverse magnetic catalysis on the thermodynamic physical quantities, we study phase diagram and some thermodynamic properties. The characteristics of the shear viscosity coefficient components of the magnetized medium system of the first order phase transition near CEP and critical endpoint transitions are investigated in Sec.III. We make summaries and conclusions in Sec.IV.

\section{NJL model in a magnetic field background}\label{sec:02 setup}

The Lagrangian density of the standard two flavors NJL model\cite{Nambu:1961tp} is as
\begin{equation}\label{eq:01}
\mathcal{L}_{\textrm{NJL}} =\overline{\psi}(i\gamma_{\mu}\partial^{\mu}-m)\psi+G[(\overline{\psi}\psi)^{2}+(\overline{\psi}i\gamma^{5}\vec{\tau}\psi)^{2}],
\end{equation}
where $\psi$  represents quark field with isospin symmetry, while $\vec{\tau}$  are isospin Pauli matrices. The Lagrangian density in Eq. (1) is invariant under global
$U(2)_{\rm f} \times SU(N_{\rm c})$ and, when $m$ = 0, the theory is also invariant under chiral $SU(2)_{\rm L} \times SU(2)_{\rm R}$. The dimensionless model coupling constant parameter $G\Lambda^{2}=2.122$~\cite{McLerran:2013hla}, the three-dimensional momentum cutoff parameter $\Lambda=650~\mathrm{MeV}$ , and the quark flow mass $m_{u}=m_{d}=5.5~ \mathrm{MeV}$ is taken. By taking the mean field approximation, one can determine the vacuum ground state meson mass $m_{\pi}=140 \textrm{MeV}$ , meson decay constant $f_{\pi}=93~\mathrm{MeV}$, and chiral condensation of the system  $\langle \bar{\psi_{f}}\psi_{f}\rangle^{1/3} = -250~\mathrm{MeV}$ , respectively.

Assuming that the magnetic field is along the $z$ axis, in the mean field approximation and for the isospin symmetric system, the thermodynamic potential of the NJL model is given by~\cite{Menezes:2008qt}
\begin{equation}\label{eq:02}
    \begin{split}
\Omega = & \frac{(M-m)^{2}}{4G} + \frac{N_{c}N_{f}}{8\pi^{2}} \{ M^{4} \ln[\frac{\Lambda+ \varepsilon_{\Lambda}}{M}] -\varepsilon_{\Lambda}\Lambda [\Lambda^{2}+\varepsilon_{\Lambda}^{2}] \}  \\
& -\frac{N_{c}}{2\pi^{2}} \sum_{f=u}^{d} (|q_{f}|B)^{2} \{ \zeta^{0,1}(-1,x_{f})-\frac{1}{2} [x_{f}^{2}-x_{f}] \ln(x_{f})+\frac{x_{f}^{2}}{4}\}  \\
& -\frac{N_{c}}{2\pi^{2}}\sum_{f=u}^{d}\sum_{k = 0}^{\infty}\alpha_{k}|q_{f}|B \int_{-\infty}^{\infty} \frac{dp_z}{2\pi} [T \ln(1+e^{-(E+\mu)/T})+ T \ln(1+e^{-(E-\mu)/T})] . \end{split}
\end{equation}

The quark self-energy caused by a four-fermion interaction produces
the constituent quark mass $M$ in the \textrm{NJL} model.  The constituent quark mass $M$
can be obtained from the gap equation \cite{Buballa:2003qv,Menezes:2008qt,Ferrari:2012yw} as
\begin{equation}\label{eq:03}
 M = m - 2G  \sum_{f=u}^{d}\langle \bar{\psi_{f}}\psi_{f}\rangle.
\end{equation}

The constituent quark mass $M$ at a finite temperature $T$ and a strong magnetic field $eB$  is given by
\begin{equation}\label{eq:04}
    \begin{split}
 M = & m + 2G\frac{N_{c}N_{f} M}{2\pi^{2}} \{\Lambda \varepsilon_{\Lambda} -M^{2}\ln[\frac{\Lambda+ \varepsilon_{\Lambda} }{M}]\}  \\
   & +2G\frac{N_{c}M}{2\pi^{2}}\sum_{f=u}^{d}|q_{f}|B \{\ln[\Gamma(x_{f})]-\frac{1}{2}\ln(2\pi)+x_{f}-\frac{1}{2}(2x_{f}-1)\ln(x_{f})\} , \\
   & -2G\frac{N_{c}M}{4\pi^{2}} \sum_{f=u}^{d}\sum_{k=0}^{\infty}\alpha_{k}|q_{f}|B \int_{-\infty}^{\infty}\frac{dp_z}{E}\{\frac{1}{e^{(E-\mu)/T}+1}+\frac{1}{e^{(E+\mu)/T}+1}\} \end{split}
\end{equation}
where color freedom is $N_{c}=3$, flavor freedom is $N_{f}=2$, $\varepsilon_{\Lambda}=\sqrt{\Lambda^{2}+M^{2}}$ is the effective cut energy, $q_{f}$ is the quark electric charge ( $|q_{u}|=\frac{2}{3}e$, $|q_{d}|=\frac{1}{3}e$, $e=\frac{1}{\sqrt{137}}$), $\zeta$ is the Riemann-Hurwitz zeta function, $\alpha_{k}=2-\delta_{0k}$  is the degeneracy factor, where   $k$ is related to the Landau energy level and spin value, and the quark dispersion relationship in the magnetic field is $E=\sqrt{p_{z}^{2}+2k|q_{f}|B+M^{2}}$  and $x_{f}=\frac{M^{2}}{2|q_{f}|B}$. Since the mean field approximation is adopted with $m = m_{u} = m_{d}= 5.5~\mathrm{MeV}$, there exists a dynamic mass relation $M = M_{u}= M_{d}$ for $u$ and $d$ quarks.

As previously mentioned, the NJL model in the quasiparticle approximation cannot describe inverse magnetic catalysis (IMC) of lattice simulation~\cite{Bali:2011qj,Bali:2012zg}, unless it is specified that the coupling constant $G$ of the model depends on $T$ and $B$ \cite{Farias:2014eca,Farias:2016gmy}. Further evidence that the correlation between IMC phenomenon and magnetic field reduces the coupling strength of effective degrees of freedom of quark matter was given in the Refs.\cite{Ayala:2014iba, Ayala:2014gwa}.

Under the condition of strong magnetic field $\sqrt{eB}\gg\Lambda_{\textrm{QCD}}$ , the first term of the expansion of the running coupling constant $\alpha_{\textrm{s}}$  of QCD with the magnetic field  $eB$  is given \cite{Miransky:2002rp} as
\begin{equation}\label{eq:04}
\frac{1}{\alpha_{s}}\sim b\ln\frac{eB}{\Lambda_{QCD}^{2}} ,
\end{equation}
where $b=\frac{11N_{c}-2N_{f}}{12\pi}$. Therefore, in the article we make use of an ansatz that makes $G$ a running coupling with $eB$ and $T$ as $G(B,T)$ , very much like the strong-coupling runs in QCD. As a result, one can find that , the model achieves magnetic catalysis at $T=0$, but it reaches inverse magnetic catalysis near the phase transition point which is in qualitative agreement with the lattice simulations of Refs.\cite{Bali:2011qj,Bali:2012zg}.
\begin{equation}\label{eq:05}
G(B,T)=\frac{G_{0}}{\ln(e+\beta(B)(T/\Lambda_{\textrm{QCD}})^{6})} ,
\end{equation}
where $\Lambda_{\textrm{QCD}}$  is the QCD energy scale, and the induced parameter $\beta(B)$  is assumed to be dependent  on the magnetic field.

\begin{figure}
    \centering
    \includegraphics[width=8.5cm]{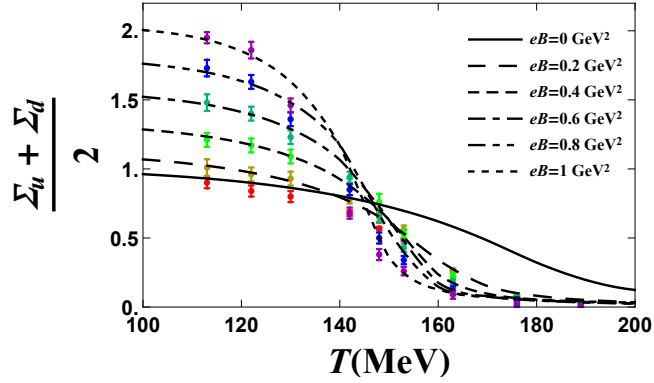}
    \caption{\label{fig1} Model fitting results after coupling constant correction $G(B,T)$  under different magnetic fields. The error bar in the figure represents the quark condensation correlation value given by the lattice QCD \cite{Bali:2011qj,Bali:2012zg}, and the line represents the calculation result of our modified model.}
\end{figure}

\begin{table}
\caption{\label{tab:1}Values of parameters under different magnetic fields. The unit of magnetic field $eB$ is $\mathrm{GeV}^{2}$. In addition, in order to compare with the original NJL model, we make the coupling constant $G$ consistent with $G_{0}$  in the case of $eB=0$.}
\begin{ruledtabular}
\begin{tabular}{lllllll}
 $eB (\mathrm{GeV}^{2}) $  & 0.0 & 0.2  & 0.4  & 0.6  & 0.8  & 1.0  \\
 \hline
 $\beta(B)$  & 0.0 & 1.3  & 3.6  & 8.0  & 16.0  & 32.0   \\
\end{tabular}
\end{ruledtabular}
\end{table}

By fitting with the LQCD results with the improved \textrm{NJL} model, we study the dependent of $(\Sigma_{u}+\Sigma_{d})/2$ on temperature $T$ as shown in Fig. 1, and get the fitting parameter  $\beta(B)$ of $G(B,T)$  from Eq.(5) as shown in Table 1. $\Sigma_{f}=\Sigma_{f}(B,T)$ ($f = u, d$ ) shown in Fig. 1  is defined as
\begin{equation}\label{eq:06}
\Sigma_{f}(B,T)=\frac{2m_{f}}{m_{\pi}^{2}f_{\pi}^{2}}[\langle\bar{\psi_{f}}\psi_{f}\rangle (B,T)-\langle\bar{\psi_{f}}\psi_{f}\rangle (0,0)]+1.
\end{equation}

It can be seen from Fig. 1 that at a lower temperature, the chiral quark condensation increases with the magnetic field, but drops rapidly near the phase transition temperature.
When $T\geq T_{c}$, chiral quark condensation decreases with magnetic field, and it is characterized by inverse magnetic catalysis(IMC).

In this work we discuss the physical point with nonzero current quark masses so that, at high temperatures, the phase diagram manifests a crossover where chiral symmetry is partially restored. In this case, one can only establish a pseudocritical temperature $T_{pc}$ which depends on the observable used to define it.
Fig. 2 is not directly obtained by using the running coupling given by Eq. 6. It is obtained from the gap equation of Eq.(3) at first, and then use the location of the peaks for the thermal susceptibility $-\frac{\textrm{d} \langle\bar{\psi_{f}}\psi_{f}\rangle}{\textrm{d} T}$  to determine the dependence of phase transition temperature on the magnetic field.

\begin{figure}
    \centering
    \includegraphics[width=7.5cm]{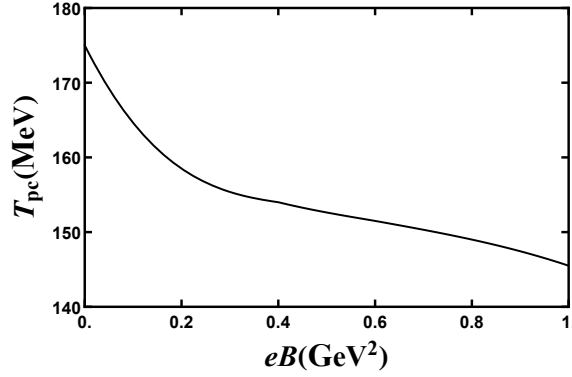} \vskip0.2cm
    \caption{\label{fig2} The dependence of pseudocritical temperature $T_{pc}$ on magnetic field at zero chemical potential. The unit of magnetic field strength $eB$  is $\mathrm{GeV}^{2}$, and the unit of pseudocritical temperature $T_{pc}$  is $\mathrm{MeV}$.}
\end{figure}

The dependence of the pseudo-critical temperature $T_{pc}$  on the magnetic field $eB$ is given shown in Fig. 2. It can be seen that the modified \textrm{NJL} model has obvious \textrm{IMC} effect, that is, with the increase of magnetic field, the pseudo-critical temperature $T_{pc}$ gradually decreases with magnetic field.

After considering strong magnetic field at zero chemical potential, one can extend it to the case of finite chemical potential. In the first order phase transition region, the phase diagram $T-\mu$  has an obvious phase transition boundary, and the phase transition temperature is computed as that the thermodynamic potential of the system has two degenerate minimum points. The \textrm{CEP} is obtained from the condition of the zero-curvature of an effective potential or equivalently from the divergence of the quark-number susceptibility.
\begin{figure}
    \centering
    \includegraphics[width=8.5cm]{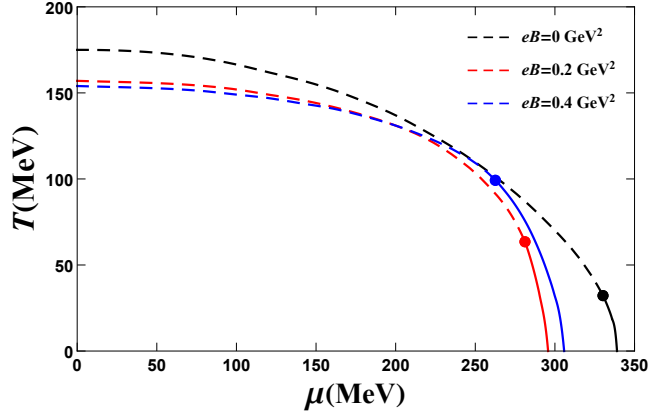}
    \caption{\label{fig3} Phase diagrams in the presence of strong magnetic fields $eB = 0.0, 0.2$ and $0.4 ~\mathrm{GeV}^{2}$ respectively.
The solid lines correspond to first-order transition, the dash dot lines correspond to crossover phase transition, and
the full dots correspond to CEP.}
 \end{figure}

Figure 3 we publish the phase diagram with magnetic fields of $eB=0.0~\mathrm{GeV}^{2}$, $eB=0.2~\mathrm{GeV}^{2}$, and $eB=0.4~\mathrm{GeV}^{2}$, respectively. It is found that the crossover occurs at high temperature and small chemical potentials $\mu$ while the first-order phase transition happens at low temperatures $T$ and large chemical potential $\mu$.  When the magnetic field rises from $0.0$ to $0.4 ~\mathrm{GeV}^{2}$, the first order lines are shortened first and then become longer as the magnetic fields become stronger. The results are consistent with those of Ref.~\cite{Su:2021cda}.

The phase transition between the crossover and first order transition is separated by a \textrm{CEP} in $T-\mu$  plane. The existence of \textrm{CEP} is an important topic in \textrm{QCD} phase diagram and the location of CEP has been investigated by lattice QCD and some effective models \cite{Fodor:2004nz, Costa:2015bza}. Figure 3 displays the locations of CEP at $eB=0.0~\mathrm{GeV}^{2}$, $eB=0.2~\mathrm{GeV}^{2}$, and $eB=0.4~\mathrm{GeV}^{2}$, respectively. When $eB=0.0~\mathrm{GeV}^{2}$, $eB=0.2~\mathrm{GeV}^{2} $, and $eB=0.4~\mathrm{GeV}^{2}$, the \textrm{CEP} is located at ($T$, $\mu$)=(32.2, 330.2) MeV,  ($T$, $\mu$) = (63.5, 281.2) MeV, and  ($T$, $\mu$)= (99.2, 262.6) MeV, respectively. It is found that the temperature of  \textrm{CEP} will increase with the magnetic field. When the magnetic fields become stronger, the chemical potential of the \textrm{CEP} decreases with the temperature of~\textrm{CEP}.

Based on the thermodynamic potential, the entropy density $s$ and net particle number density $n$ of the system is given as
\begin{equation}\label{eq:07}
n=-\frac{\partial \Omega}{\partial \mu} {\Big |}_{T} ,
\end{equation}
and
\begin{equation}\label{eq:08}
s=-\frac{\partial \Omega}{\partial T} {\Big |}_{\mu}.
\end{equation}

The entropy density $s$ and net particle number density $n$ of the system with the strong magnetic field are
\begin{equation}\label{eq:09}
\begin{split}
 s = & \sum_{f = u}^{d}\sum_{k=0}^{\infty}\alpha_{k}\frac{N_{c}|q_{f}|B}{4\pi^{2}} \int_{-\infty}^{\infty}dp_{z} \{ \ln(1+e^{-(E+\mu)/T})  \\
     &  +\frac{E+\mu}{T} \frac{1}{1+e^{(E+\mu)/T}}+\ln(1+e^{-(E-\mu)/T})+\frac{E-\mu}{T} \frac{1}{1+e^{(E-\mu)/T}} \},
 \end{split}
\end{equation}
and
\begin{gather}\label{eq:010}
 n = \sum_{f=u}^{d}\sum_{k=0}^{\infty}\alpha_{k}\frac{N_{c}|q_{f}|B}{4\pi^{2}} \int_{-\infty}^{\infty}dp_{z} \{ \frac{1}{1+e^{(E-\mu)/T}}- \frac{1}{1+e^{(E+\mu)/T}} \}.
\end{gather}

\begin{figure}
    \centering
    \includegraphics[width=8.5cm]{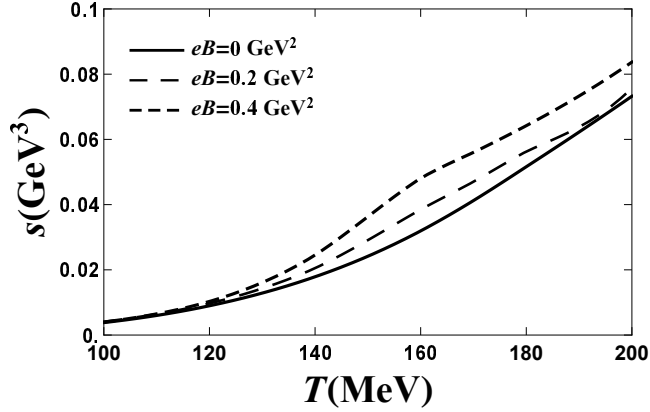}
    \caption{\label{fig4} The dependence of entropy density of \textrm{QCD} medium on temperature with different magnetic field at chemical potential $\mu =0 $. The unit of entropy
 density is $\mathrm{GeV}^{3}$.}
\end{figure}

The dependence of entropy density on temperature with different magnetic fields is displayed in Fig. 4. At a lower temperature ($T < T_{pc}$), entropy density almost does not change with magnetic field, but when the temperature rises to the $T_{pc}< T\leq 180~\mathrm{MeV}$ region, the greater the magnetic field is, the faster the entropy increasing with the temperature is.

\section{Shear viscosity coefficient in strong magnetic field}\label{sec:03 contents}
\subsection{Theory of shear viscosity in the presence of strong magnetic field}
By assuming the thermal system is near equilibrium, one can use the Boltzmann equation of the kinetic theory to calculate the dependence of the shear viscosity on thermal parameters. The relaxation time of quarks is obtained from the thermally averaged total cross section of elastic scattering in the dilute gas approximation. We expand the distribution function of the nonideal fluid system near the equilibrium state as
\begin{equation}\label{eq:011}
f(x,p)=f^{0}(x,p)+f^{1}(x,p),
\end{equation}
where $f^{0}(x,p)$  is the equilibrium distribution function, which is written as covariant form
\begin{equation}\label{eq:012}
f^{0}(x,p)= \frac{1}{exp[\frac{u_{\mu}(x)p^{\mu}\mp \mu(x)}{T(x)}]+1},
\end{equation}
where $u_{\mu}(x)=\gamma_{\mu}(1,\textbf{u}(x))$ is the four velocities of the fluid. Under the relaxation time approximation, one writes the covariant Boltzmann equation of motion as
\begin{equation}\label{eq:013}
\frac{df}{dt}=\frac{p^{\mu}}{E}\partial_{\mu}f-\frac{\partial {E}}{\partial {x^{i}}} \frac{\partial {f}}{\partial {p^{i}}}=-\frac{f^{1}}{\tau},
\end{equation}
where $\tau$  is the relaxation time. In a dissipative fluid, the dependence among the energy dynamic tensors of the fluid on the distribution function is given as
\begin{equation}\label{eq:014}
\delta T^{ij} = \sum_{a}g_{a}\int \frac{d^{3}p}{(2\pi)^{3}}\frac{p^{i}p^{j}}{E_{a}} \delta \tilde{f_{a}} ,
\end{equation}
where $g_{a}$  is the degeneracy of particle type $a$ , and there is always $\tilde{f}=f^{1}$  in the above equation. From Newton's viscosity law, one can take
\begin{equation}\label{eq:015}
\delta T^{ij} = -\zeta\delta_{ij}\partial_{k}u^{k}-\eta \omega_{ij} ,
\end{equation}
where $\eta$ is the shear viscosity coefficient of viscous fluid, $\zeta$ is the bulk viscosity coefficient, and $\omega_{ij}$ is the partial derivative of flow velocity. Comparing the coefficients from the dissipative part of the energy-momentum tensor \cite{Ghosh:2018cxb}, one can obtain the  shear viscosity coefficient
\begin{equation}\label{eq:016}
\eta = \frac{1}{15T}\sum_{a}g_{a}\int \frac{d^{3}p}{(2\pi)^{3}}\frac{p^{4}}{E_{a}^{2}}(\tau_{a}f_{a}^{0}(1-f_{a}^{0})) .
\end{equation}

In the presence of an external magnetic field, the shear viscosity coefficients of the magnetized medium will have five components, of which $\eta_{0}$  is not affected by the magnetic field, and the remaining four components are related to the magnetic field \cite{Ghosh:2018cxb}.

In order to calculate the collision relaxation time of the system, the elastic scattering theory of the two-body collision $ a + b\rightarrow c+d $ is adopted in our study. Near the phase transition temperature, the constituent particles of the system medium are mainly $u$ and $d$ quarks and their antiquarks, so there are a total of $12$ collision scenarios as
\begin{equation}\label{eq:017}
\begin{split}
 & uu\rightarrow uu,  ud\rightarrow ud,  u\bar{u}\rightarrow u\bar{u},  u\bar{u}\rightarrow d \bar{d},  \\
 & u\bar{d}\rightarrow u\bar{d},  dd\rightarrow dd, \bar{u}d\rightarrow \bar{u}d,  d\bar{d}\rightarrow d \bar{d},  \\
 & d\bar{d}\rightarrow u\bar{u},  \bar{u}\bar{u}\rightarrow \bar{u}\bar{u},  \bar{u}\bar{d}\rightarrow \bar{u}\bar{d},  \bar{d}\bar{d}\rightarrow \bar{d}\bar{d}.
\end{split}
\end{equation}

If the incident particle $a$ is used as a probe, one can evaluate its relaxation time by $\tau_{a}^{c}=\frac{1}{\Gamma_{a}^{c}}$ , and the collision width $\Gamma_{a}^{c}$  is given as averaging its collision width with different momentum
\begin{equation}\label{eq:018}
\Gamma_{a}^{c} = \frac{\int \frac{d^{3}p_{a}}{(2\pi)^{3}} \Gamma_{a}^{c}(p_{a})f_{a}}{\int \frac{d^{3}p_{a}}{(2\pi)^{3}}f_{a}},
\end{equation}
where $\Gamma_{a}^{c}(p_{a})$ is
\begin{equation}\label{eq:019}
\Gamma_{a}^{c}(p_{a}) = \sum_{a} \int \frac{d^{3}p_{b}}{(2\pi)^{3}} \sigma_{ab}(p_{a},p_{b}) \nu_{ab}(p_{a},p_{b})f_{b},
\end{equation}
and $\sigma_{ab}$  is the collision cross section and $\nu_{ab}$  is the relative velocity of two particles, which are respectively given as
\begin{equation}\label{eq:020}
\sigma_{ab}(p_{a},p_{b})=\frac{1}{16\pi s}|\overline{M}_{ab}|^{2} ,
\end{equation}
and
\begin{equation}\label{eq:021}
\nu_{ab}(p_{a},p_{b})=\frac{(E_{a}+E_{b})\sqrt{(E_{a}+E_{b})^{2}-4M^{2}}}{2E_{a}E_{b}},
\end{equation}
where $s=(E_{a}+E_{b})^{2}$  is Mandelstam variable, and scattering matrix $|\overline{M}_{ab}|^{2}$ is taken as
\begin{equation}\label{eq:022}
|\overline{M}_{ab}|^{2}=\frac{1}{2\times 2} G^{2}\times 16(\frac{s}{2})^{2}=G^{2}s^{2}.
\end{equation}

If the particle $a$ is in a strong magnetic field environment, the relaxation time generated by the magnetic field is $\tau_{a}^{B}=\frac{E_{a}}{\mid q_{a}\mid B}$ . Most of the particles are bound at the lowest Landau energy level in the strong magnetic field background, and the ground state particle number density of whole energy levels still dominates. The larger the magnetic field is, the higher the proportion of the lowest Landau level particles is.

The dependence of the collision relaxation time $\tau_{c}$ on temperature with magnetic field at chemical potential $\mu=0$ is displayed in Fig. 5. The relaxation time $\tau_{c}$ decreases rapidly with the temperature, especially at low temperature. One also finds that at the same temperature, the larger the magnetic field is, the larger the relaxation time is. It is found that when the temperature is near the phase transition temperature $150-180~\mathrm{MeV}$, the corresponding relaxation time is between $10~\mathrm{fm}$ and $20~\mathrm{fm}$.

\begin{figure*}
    \centering
    \includegraphics[width=8.5cm]{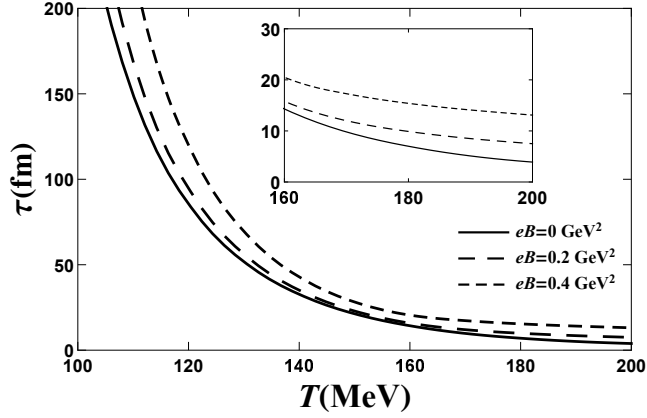}
    \caption{\label{fig5} The dependence of the collision relaxation time on temperature in the magnetic
field environment at chemical potential $\mu=0$. The unit of relaxation time $\tau$ is $\mathrm{fm}$.}
\end{figure*}

In the presence of an strong magnetic field, the shear viscosity coefficient is split into five different components because of anisotropy in tangential stress of the fluid. When the magnetic field is greater than $10m_{\pi}^{2}$ ($\approx 0.2~\mathrm{GeV}^{2}$), $\tau^{c}\gg \tau^{B} $  is satisfied, so we can use the strong field limit to study shear viscosity of phase transition. $\eta_{0}$  is not influenced by the magnetic field, but the other four of five components can be merged to two components $\eta_{2}$  and $\eta_{4}$  in the strong field limit \cite{Tuchin:2011jw}.

Thus, in the presence of a strong magnetic field, the expressions of the four components of the shear viscosity for the quark are
\begin{equation}\label{eq:023}
\eta_{2}=4\eta_{1}=\frac{1}{15T}\sum_{a} g_{a} \int \frac{d^{3}p_{a}}{(2\pi)^{3}} \frac{p^{4}}{E_{a}^{2}} f_{a}^{0}(1-f_{a}^{0}) \frac{(\tau_{a}^{B})^{2}}{\tau_{a}^{c}},
\end{equation}
and
\begin{equation}\label{eq:024}
\eta_{4}=2\eta_{3}=\frac{1}{15T}\sum_{a} g_{a} \int \frac{d^{3}p_{a}}{(2\pi)^{3}} \frac{p^{4}}{E_{a}^{2}} f_{a}^{0}(1-f_{a}^{0}) \tau_{a}^{B},
\end{equation}
where $\tau_{a}^{c}$  is the relaxation time generated by the collision mechanism, and $\tau_{a}^{B}$ is the relaxation time induced by the magnetic field. It can be seen from Eqs. (23) and (24) that $\eta_{2}$  depends not only on $\tau_{a}^{c}$, but also on $\tau_{a}^{B}$, and $\eta_{4}$  only depends on $\tau_{a}^{B}$  with the strong field limit. In the following, we will use the strong field limit  to study shear viscosities near chiral phase transition of the magnetized QCD medium.

\subsection{The computed results of the shear viscosities}
 The dependencies of shear viscosity coefficient and the ratio of shear viscosity coefficient to entropy density on temperature at $\mu=0$ and $eB = 0$ are displayed in Fig. 6. At zero chemical potential and zero magnetic fields, the viscosity coefficient $\eta$  decreases with the increase of temperature. At the same time, one also finds that the ratio of shear viscosity coefficient to entropy density $\eta/s$ decreases with temperature, but the rate of decrease is obviously faster than the viscosity coefficient, mainly because the entropy density value is large.
\begin{figure}
    \centering
    \includegraphics[width=8.5cm]{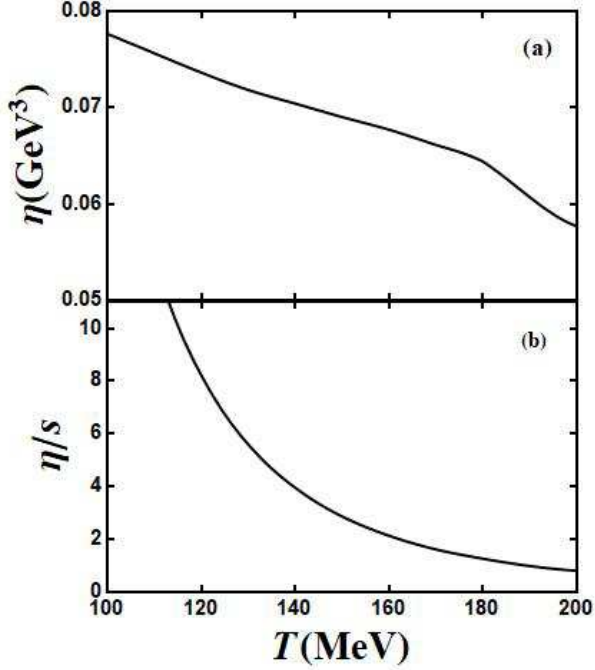}
    \caption{\label{fig6} The dependencies of shear viscosity coefficient and the ratio of shear viscosity coefficient to entropy density change on temperature at $\mu=0$ and $eB = 0$. (a) Shear viscosity coefficient $\eta$, the unit of shear viscosity coefficient $\mathrm{GeV}^{3}$ ; (b) The ratio of shear viscosity coefficient to entropy density.}
\end{figure}

\begin{figure}
    \centering
    \includegraphics[width=8.5cm]{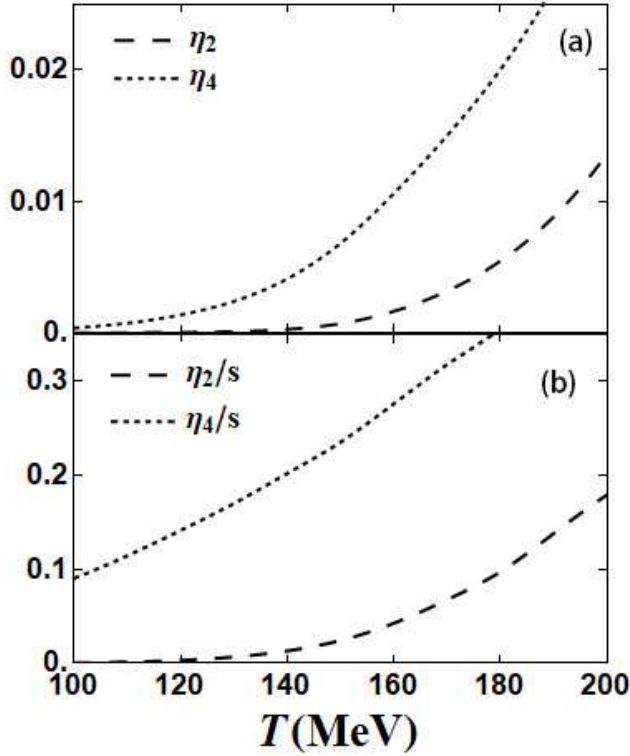}
    \caption{\label{fig7} The dependencies of the two components of $\eta_{2}$ and $\eta_{4}$ of the shear viscosity and the ratio of the two components of the shear viscosity to entropy density on temperature with $eB = 0.2~\mathrm{GeV}^{2}$  at $\mu = 0$. (a) For $\eta_{2}$ and $\eta_{4}$  of the shear viscosity with temperature; (b) For the ratio of the two components of the shear viscosity to entropy density.}
    \end{figure}

\begin{figure}
    \centering
    \includegraphics[width=8.5cm]{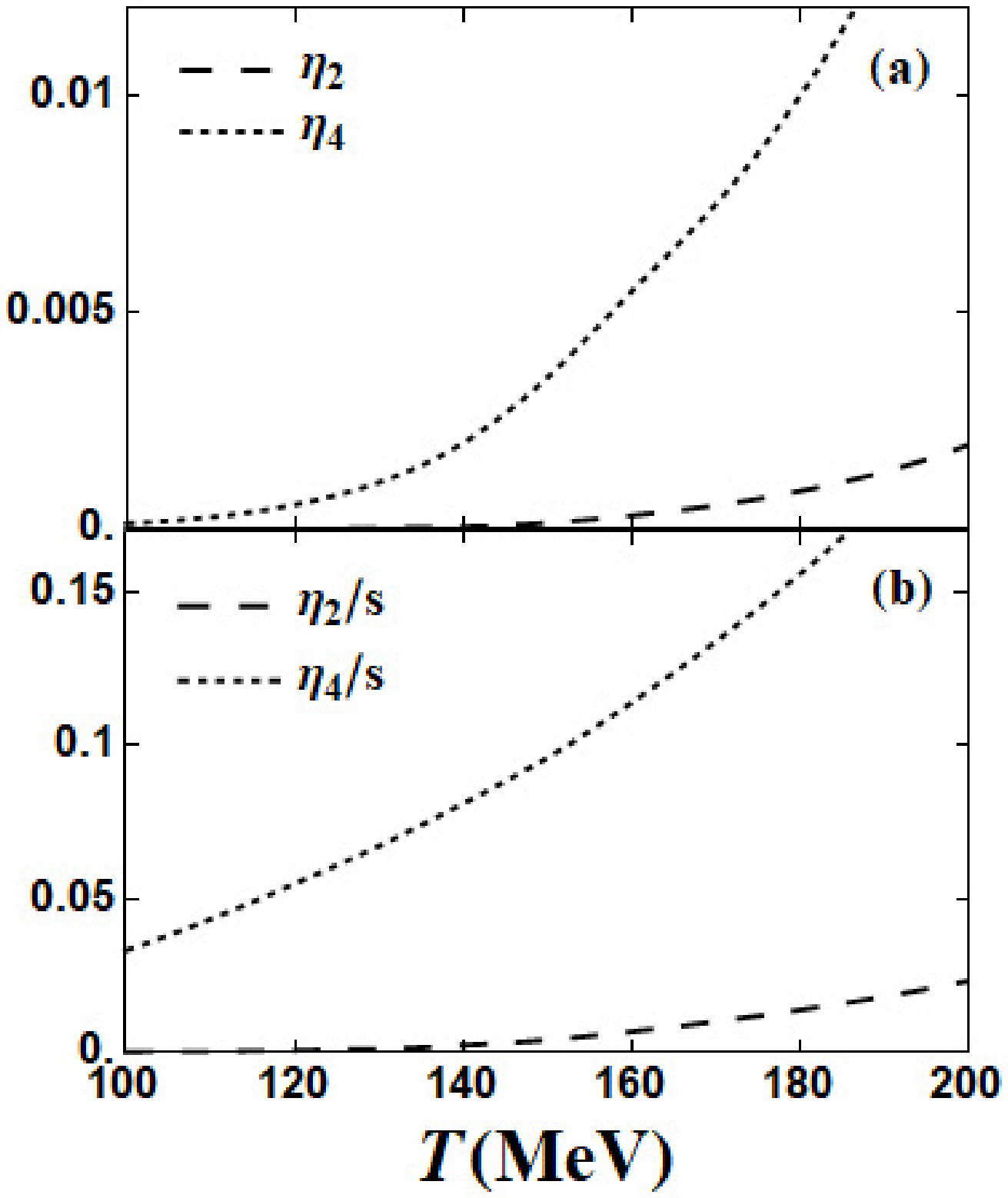}
    \caption{\label{fig8} The same as Fig. 7 but for $eB = 0.4~\mathrm{GeV}^{2}$.}
\end{figure}

Comparing with the dependence of $\eta$ with temperature for zero magnetic field shown in Fig. 6, one finds that the components $\eta_{2}$ and $\eta_{4}$ of shear viscosity  with magnetic field shown in Figs. 7 and 8 both are increasing functions of temperature. From Fig. 5, one finds that the collision relaxation time $\tau^{c}$  is a rapidly decreasing function of temperature. However, $\tau^{c}$  appears on the denominator in Eqs.(23) and (24), which makes the effective relaxation time rise overall. Therefore,$\eta_{2}$ and $\eta_{4}$ increase with temperature mainly due to the change of effective relaxation time when magnetic field is applied. By comparing Fig. 7 with Fig. 8, one can find that  $\eta_{2}$ and $\eta_{4}$  both decrease with the magnetic field.

\begin{figure}
    \centering
    \includegraphics[width=8.5cm]{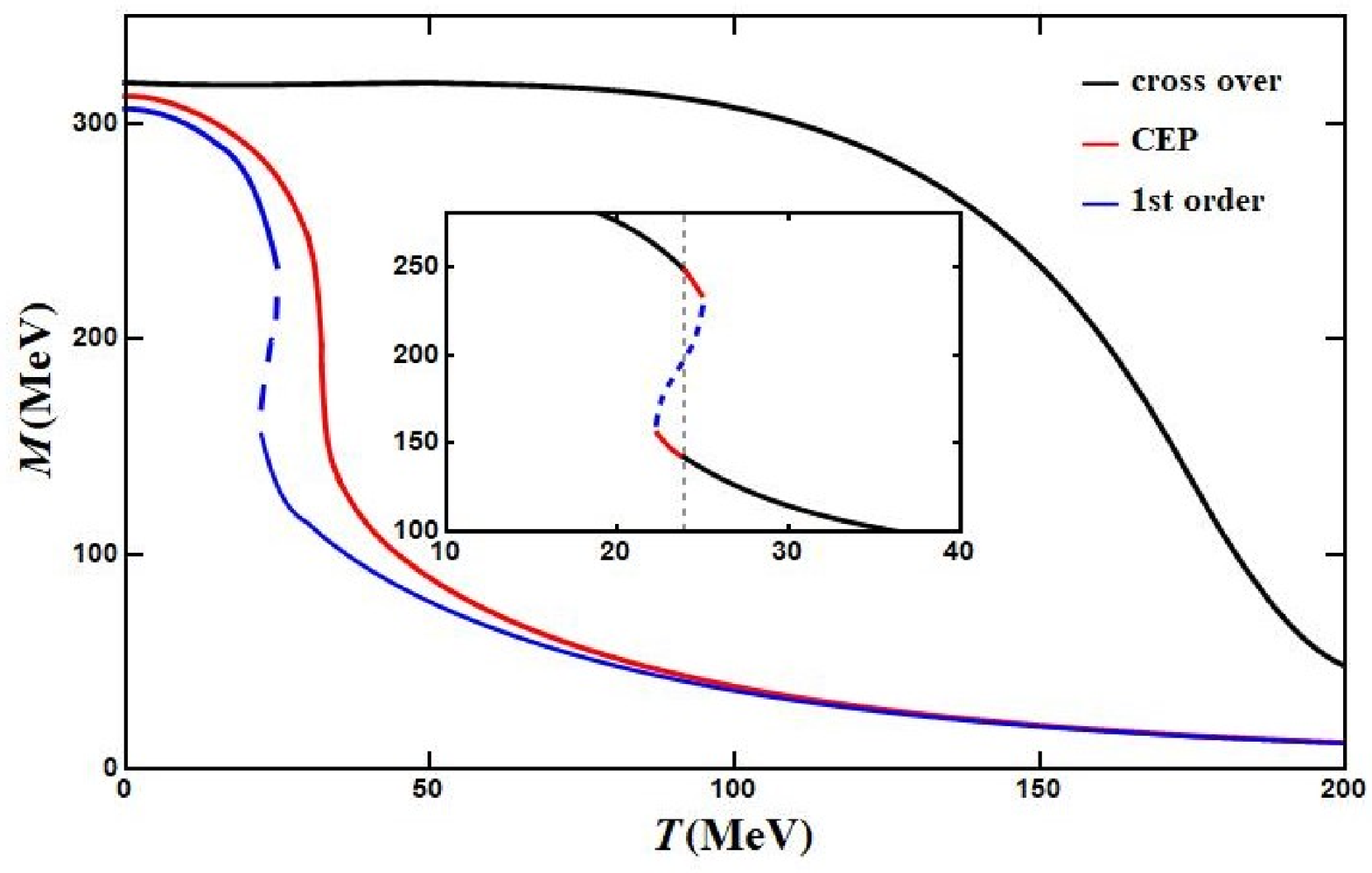}
    \caption{\label{fig9} The dependencies of the dynamic quark mass $M$ on temperature of the crossover, the first order phase transition and the \textrm{CEP} transition, respectively. The black solid line corresponds to the crossover phase transition, the red solid line corresponds to the \textrm{CEP} phase transition and the blue solid line corresponds to the first-order phase transition.  The detailed discontinuous area part of the dynamic quark mass $M$ with temperature $T$ ($M-T$) of the
    first order phase transition is shown in the small box diagram inside Fig. 9.}
\end{figure}

Figure 9 shows the dependence of the dynamic quark mass $M$ on temperature of the crossover phase transition, the first order phase transition and the critical endpoint (CEP) phase transition respectively. The first order phase transition, which is the phase transition from the hadron phase to the quark gluon plasma phase, has obvious phase boundary.
For the first order phase transition, its dynamic mass $M$ has a discontinuous change at the phase transition point, which is called abrupt change. The thermodynamic potential $\Omega = 0$  of the first order phase transition has three extreme points, corresponding to two minima and one maximum of the partial derivative (by $\frac{\partial \Omega}{\partial M}$ ), respectively. Obviously, the two minima of thermodynamic potential $\Omega$  are the physical results that meet the requirements, corresponding to the thermodynamic potential of the state of the quark phase and the hadron phase, respectively. The maximum of the thermodynamic potential $\Omega$  must be eliminated by $\frac{\partial^{2} \Omega}{\partial M^{2}}>0$.

The stable solution is the one which corresponds to the global minimum of the thermodynamic potential, $\Omega$. The other solutions, marked by the red lines and blue dashed lines in the inset plot of Fig. 9, represent the metastable and the unstable solutions which respectively correspond to the local minimum and the maximum of $\Omega$.
For the crossover phase transition, the thermodynamic potential of the system has only one extreme point, which corresponding to that the dynamic mass varies continuously with temperature. The so-called \textrm{CEP} phase transition is the junction point between the crossover phase transition and the first order phase transition on the phase diagram, which is also the endpoint of the first order phase transition, so it is called the \textrm{CEP}.

\begin{figure}
    \centering
    \includegraphics[width=8.5cm]{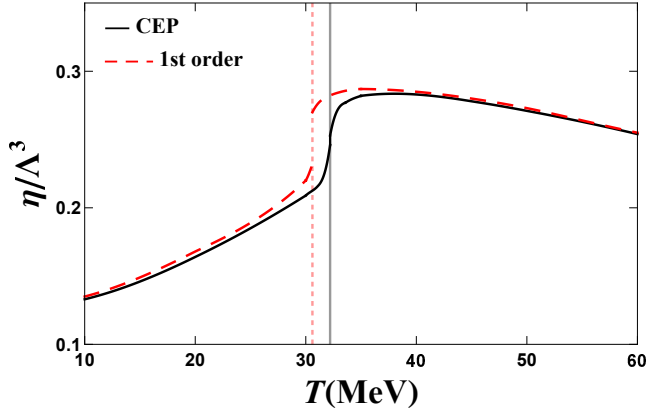}
    \caption{\label{fig10} The shear viscosity normalized by the 3-momentum cutoff along the phase boundary ($\Lambda=650~\mathrm{MeV}^{3}$ ). The vertical dotted-line (black line) indicates the location of the \textrm{CEP}, and the vertical dotted-line (red line) indicates the location of the first order phase transition point.}
\end{figure}

 The renormalized shear viscosity ($\eta/\Lambda^{3}$) along the chiral phase boundary obtained from the modified \textrm{NJL} model at $eB = 0$ is shown Fig. 10. To study the dependencies of shear viscosity coefficient with temperature for the CEP transition, we fix the chemical potential $\mu=\mu_{\textrm{CEP}}$. For the first order phase transition point, we choose ($T_{c}, \mu_{c}) = (30.6, 331)~~\mathrm{MeV}$ near the critical endpoint. For the first order phase transition, $\eta/\Lambda^{3}$  rises with the temperature, and reaches the first order phase transition point ($T_{c}$,$\mu_{c}$)=$(31, 334)~\mathrm{MeV}$. $\eta/\Lambda^{3}$  has a discontinuous at the phase transition point ($T_{c}, \mu_{c}$), and $\eta/\Lambda^{3}$ exists an upward jump from hadron phase to QGP phase. After the phase transition, $\eta/\Lambda^{3}$ will soon reach the maximum value, and then begin to decrease with temperature. For the CEP transition, the situation is very similar to that of the first order phase transition. From low temperature, $\eta/\Lambda^{3}$  rises with temperature to reach at the CEP position. The only difference is that $\eta/\Lambda^{3}$ of CEP at the phase transition point is a continuous change.

\begin{figure}
    \centering
    \includegraphics[width=8.5cm]{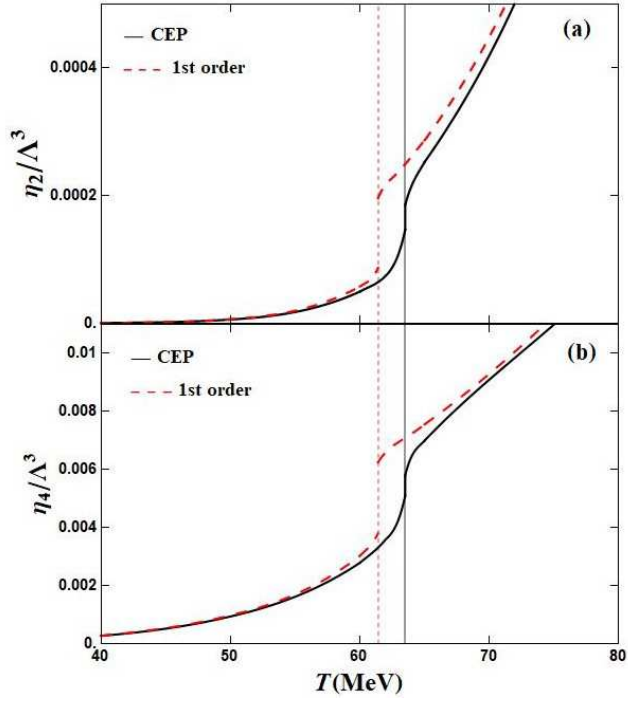}
    \caption{\label{fig11} The dependencies of renormalized shear viscosity coefficient components ($\eta_{2}$ and $\eta_{4}$ ) on temperature of CEP phase transition and first order phase transition respectively with magnetic field  ($eB = 0.2 \mathrm{GeV}^{2}$). The vertical dotted-line (black line) indicates the position of the CEP, and the vertical solid line (read line) indicates the position of the first order phase transition point. (a) For $\eta_{2}/\Lambda^{3}$, and (b) For $\eta_{4}/\Lambda^{3}$.}
\end{figure}

The dependencies of renormalized shear viscosity coefficient components ($\eta_{2}$ and $\eta_{4}$) on temperature of CEP phase transition, and first order phase transition with $eB = 0.2~\mathrm{GeV}^{2}$ are shown in Fig. 11. We fix the chemical potential $\mu=\mu_{\textrm{CEP}}$ to study the shear viscosity coefficient components for the CEP transition. For the first order phase transition point, we also fix $\mu=\mu_{\textrm{C}}$ near the critical endpoint.  For the first order phase transition, starting from low temperature, $\eta_{2}/\Lambda^{3}$ and $\eta_{4}/\Lambda^{3}$ both rise with the temperature, and then reach the first order phase transition point. Each of $\eta_{2}/\Lambda^{3}$ and $\eta_{4}/\Lambda^{3}$ has a discontinuous at the phase transition point, an upward jump from hadron phase to QGP phase is shown in Fig. 11. After the first order phase transition, $\eta_{2}/\Lambda^{3}$ and $\eta_{4}/\Lambda^{3}$ both continue to increase with temperature. For the CEP transition, the dependencies of renormalized shear viscosity coefficient components on temperature are very similar to that of the first order phase transition. Unlike the first order phase transition, $\eta_{2}/\Lambda^{3}$ and $\eta_{4}/\Lambda^{3}$  of CEP at the phase transition point are continuous change.

\begin{figure}
    \centering
    \includegraphics[width=8.5cm]{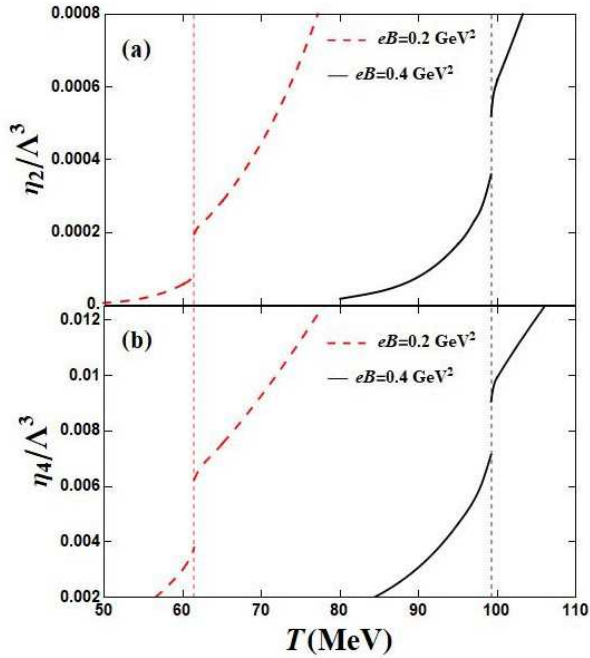}
    \caption{\label{fig12} The dependencies of renormalized shear viscosity coefficient components ($\eta_{2}$ and $\eta_{4}$ ) on temperature of first order phase transition with different magnetic fields $eB = 0.2~\mathrm{GeV}^{2}$ (solid line) and $eB = 0.4~\mathrm{GeV}^{2}$ (dashed line). The three-dimensional momentum cutoff parameter ($\Lambda=650~\mathrm{MeV}^{3}$). (a) for $\eta_{2}/\Lambda^{3}$, and  (b) for $\eta_{4}/\Lambda^{3}$.}
\end{figure}

The dependencies of shear viscosity coefficient components ($\eta_{2}$ and $\eta_{4}$) on temperature of first order phase transition with different magnetic fields $eB = 0.2~ \mathrm{GeV}^{2}$ (solid line) and $eB = 0.4~\mathrm{GeV}^{2}$ (dashed line) are shown in Fig. 12. The following two characteristics have been displayed as follows: (1) Discontinuities of $\eta_{2}/\Lambda^{3}$ and $\eta_{4}/\Lambda^{3}$ for the first order phase transition point, and there is an upward discontinuous jump from the hadron phase of the low temperature to the QGP phase of the high  temperature; (2) $\eta_{2}/\Lambda^{3}$ and $\eta_{4}/\Lambda^{3}$, their jump values at the first order phase transition point increase with magnetic field; (3) Comparing $\eta_{2}/\Lambda^{3}$ and $\eta_{4}/\Lambda^{3}$, it is obvious that at the same temperature, $\eta_{4}/\Lambda^{3}$ is larger than $\eta_{2}/\Lambda^{3}$, and the discontinuous jump value of $\eta_{4}/\Lambda^{3}$ at the phase transition point is larger than $\eta_{2}/\Lambda^{3}$.

When there is a magnetic field, Eqs. (24) and (25) are limited by the strong field limit $\tau_{c}\gg\tau_{B}$. At this time, the relaxation time $\tau_{c}$  of the collision is no longer dominant, and the role of the magnetic field begins to dominate the interaction. Since $\eta_{4}$  is only related to $\tau_{B}$, and $\eta_{2}$  is related to $\tau_{B}\frac{\tau_{B}}{\tau_{c}}$, which is equivalent to multiplying a small quantity $\frac{\tau_{B}}{\tau_{c}}$  on $\tau_{B}$, therefore, $\eta_{2}$ is always smaller than  $\eta_{4}$.

With the increase of magnetic field $eB$, due to the existence of inverse magnetic catalytic effect near the phase transition temperature, the constituent quark mass with larger magnetic field changes more sensitively and dramatically. The discontinuous jump value of constituent quark mass $M$ at the first-order phase transition point increases with the magnetic field, as shown in the Fig. 13. The corresponding discontinuous jump value of shear viscosity coefficients $\eta_{2}$  and $\eta_{4}$  at the phase transition also increases with magnetic field.

\begin{figure}
    \centering
    \includegraphics[width=8.5cm]{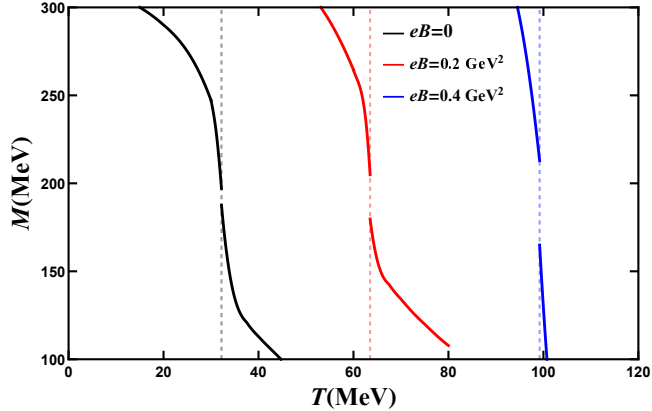}
    \caption{\label{fig13} The dependence of constituent quark mass $M$ on temperature at the first order phase transition point near CEP at $eB$ = 0.0, 0.2 and $0.4~\mathrm{GeV}^{2}$.}
\end{figure}

\section{Summary and Conclusions}\label{sec:05}

Noncentral heavy ion collisions can produce a strong magnetic field. The phase transition under such a field during the evolution of QGP to hadron gas depends mainly on transport coefficients like shear viscosity. The dissipative coefficients in the presence of magnetic field are also important and are essential ingredients for the magnetohydrodynamic evolution of the strongly interacting medium. When the magnetic field exists, the shear viscosity coefficient of the dissipative fluid system can be decomposed into five different components. In strong field limit, a detailed study of the dependencies of $\eta_{2}$  and $\eta_{4}$  on temperature and magnetic field for the first order phase transition and critical endpoint transition is studied.

A strong magnetic field introduces anisotropic feature, phase-space Landau-level quantization. It will affect the quark matter transport characteristics. Under the strong field limit, $\eta_{2}$  and $\eta_{4}$ both increase with temperature and decrease with magnetic field, and the discontinuities of $\eta_{2}$  and $\eta_{4}$ occur at the first order phase transition point. The essential reason is that the constituent quark mass of the first order phase transition is disconnected at the phase transition point, leading to the discontinuities of $\eta_{2}$  and $\eta_{4}$.

With the increase of magnetic field $eB$ , due to the existence of inverse magnetic catalytic effect near the phase transition temperature of the magnetized \textrm{QCD} medium, the constituent quark mass with larger magnetic field changes more sensitively and dramatically. The corresponding discontinuous jump value of shear viscosity coefficients  $\eta_{2}$  and $\eta_{4}$  at the phase transition also increase with the magnetic field. This implies that the shear viscosity coefficient of magnetized \textrm{QCD} medium is not a function of smooth transition in the first order phase transition temperature region.

In this paper, we only discuss the viscosity coefficient by using the two-flavors NJL model. Can we expand to the Polyakov loop (Polyakov NJL), or use the (2 + 1)- flavor NJL model to discuss the transport coefficient near the phase transition point? The answer is yes. In the future, we are planning to utilize the (2 + 1)- flavor NJL model with strong magnetic field to study the transport coefficient near the phase transition point.

\section*{Acknowledgments}
This work was supported by National Natural Science Foundation of China (Grants No. 11875178, No. 11475068,No. 11747115).

\section*{References}

\bibliography{ref}

\begin{thebibliography}{53}%
\makeatletter
\providecommand \@ifxundefined [1]{%
 \@ifx{#1\undefined}
}%
\providecommand \@ifnum [1]{%
 \ifnum #1\expandafter \@firstoftwo
 \else \expandafter \@secondoftwo
 \fi
}%
\providecommand \@ifx [1]{%
 \ifx #1\expandafter \@firstoftwo
 \else \expandafter \@secondoftwo
 \fi
}%
\providecommand \natexlab [1]{#1}%
\providecommand \enquote  [1]{``#1''}%
\providecommand \bibnamefont  [1]{#1}%
\providecommand \bibfnamefont [1]{#1}%
\providecommand \citenamefont [1]{#1}%
\providecommand \href@noop [0]{\@secondoftwo}%
\providecommand \href [0]{\begingroup \@sanitize@url \@href}%
\providecommand \@href[1]{\@@startlink{#1}\@@href}%
\providecommand \@@href[1]{\endgroup#1\@@endlink}%
\providecommand \@sanitize@url [0]{\catcode `\\12\catcode `\$12\catcode
  `\&12\catcode `\#12\catcode `\^12\catcode `\_12\catcode `\%12\relax}%
\providecommand \@@startlink[1]{}%
\providecommand \@@endlink[0]{}%
\providecommand \url  [0]{\begingroup\@sanitize@url \@url }%
\providecommand \@url [1]{\endgroup\@href {#1}{\urlprefix }}%
\providecommand \urlprefix  [0]{URL }%
\providecommand \Eprint [0]{\href }%
\providecommand \doibase [0]{http://dx.doi.org/}%
\providecommand \selectlanguage [0]{\@gobble}%
\providecommand \bibinfo  [0]{\@secondoftwo}%
\providecommand \bibfield  [0]{\@secondoftwo}%
\providecommand \translation [1]{[#1]}%
\providecommand \BibitemOpen [0]{}%
\providecommand \bibitemStop [0]{}%
\providecommand \bibitemNoStop [0]{.\EOS\space}%
\providecommand \EOS [0]{\spacefactor3000\relax}%
\providecommand \BibitemShut  [1]{\csname bibitem#1\endcsname}%
\let\auto@bib@innerbib\@empty
\bibitem [{\citenamefont {Bzdak}\ and\ \citenamefont
  {Skokov}(2012)}]{Bzdak:2011yy}%
  \BibitemOpen
  \bibfield  {author} {\bibinfo {author} {\bibfnamefont {A.}~\bibnamefont
  {Bzdak}}\ and\ \bibinfo {author} {\bibfnamefont {V.}~\bibnamefont {Skokov}},\
  }\href {\doibase 10.1016/j.physletb.2012.02.065} {\bibfield  {journal}
  {\bibinfo  {journal} {Phys. Lett. B}\ }\textbf {\bibinfo {volume} {710}},\
  \bibinfo {pages} {171} (\bibinfo {year} {2012})}\BibitemShut {NoStop}%
\bibitem [{\citenamefont {Deng}\ and\ \citenamefont
  {Huang}(2012)}]{Deng:2012pc}%
  \BibitemOpen
  \bibfield  {author} {\bibinfo {author} {\bibfnamefont {W.-T.}\ \bibnamefont
  {Deng}}\ and\ \bibinfo {author} {\bibfnamefont {X.-G.}\ \bibnamefont
  {Huang}},\ }\href {\doibase 10.1103/PhysRevC.85.044907} {\bibfield  {journal}
  {\bibinfo  {journal} {Phys. Rev. C}\ }\textbf {\bibinfo {volume} {85}},\
  \bibinfo {pages} {044907} (\bibinfo {year} {2012})}\BibitemShut {NoStop}%
\bibitem [{\citenamefont {Kharzeev}\ \emph {et~al.}(2008)\citenamefont
  {Kharzeev}, \citenamefont {McLerran},\ and\ \citenamefont
  {Warringa}}]{Kharzeev:2007jp}%
  \BibitemOpen
  \bibfield  {author} {\bibinfo {author} {\bibfnamefont {D.~E.}\ \bibnamefont
  {Kharzeev}}, \bibinfo {author} {\bibfnamefont {L.~D.}\ \bibnamefont
  {McLerran}}, \ and\ \bibinfo {author} {\bibfnamefont {H.~J.}\ \bibnamefont
  {Warringa}},\ }\href {\doibase 10.1016/j.nuclphysa.2008.02.298} {\bibfield
  {journal} {\bibinfo  {journal} {Nucl. Phys. A}\ }\textbf {\bibinfo {volume}
  {803}},\ \bibinfo {pages} {227} (\bibinfo {year} {2008})}\BibitemShut
  {NoStop}%
\bibitem [{\citenamefont {Mo}\ \emph {et~al.}(2013)\citenamefont {Mo},
  \citenamefont {Feng},\ and\ \citenamefont {Shi}}]{Mo:2013qya}%
  \BibitemOpen
  \bibfield  {author} {\bibinfo {author} {\bibfnamefont {Y.-J.}\ \bibnamefont
  {Mo}}, \bibinfo {author} {\bibfnamefont {S.-Q.}\ \bibnamefont {Feng}}, \ and\
  \bibinfo {author} {\bibfnamefont {Y.-F.}\ \bibnamefont {Shi}},\ }\href
  {\doibase 10.1103/PhysRevC.88.024901} {\bibfield  {journal} {\bibinfo
  {journal} {Phys. Rev. C}\ }\textbf {\bibinfo {volume} {88}},\ \bibinfo
  {pages} {024901} (\bibinfo {year} {2013})}\BibitemShut {NoStop}%
\bibitem [{\citenamefont {Zhong}\ \emph {et~al.}(2014)\citenamefont {Zhong},
  \citenamefont {Yang}, \citenamefont {Cai},\ and\ \citenamefont
  {Feng}}]{Zhong:2014cda}%
  \BibitemOpen
  \bibfield  {author} {\bibinfo {author} {\bibfnamefont {Y.}~\bibnamefont
  {Zhong}}, \bibinfo {author} {\bibfnamefont {C.-B.}\ \bibnamefont {Yang}},
  \bibinfo {author} {\bibfnamefont {X.}~\bibnamefont {Cai}}, \ and\ \bibinfo
  {author} {\bibfnamefont {S.-Q.}\ \bibnamefont {Feng}},\ }\href {\doibase
  10.1155/2014/193039} {\bibfield  {journal} {\bibinfo  {journal} {Adv. High
  Energy Phys.}\ }\textbf {\bibinfo {volume} {2014}},\ \bibinfo {pages}
  {193039} (\bibinfo {year} {2014})}\BibitemShut {NoStop}%
\bibitem [{\citenamefont {Feng}\ \emph {et~al.}(2018)\citenamefont {Feng},
  \citenamefont {Pei}, \citenamefont {Sun}, \citenamefont {Zhong},\ and\
  \citenamefont {Yin}}]{Feng:2016srp}%
  \BibitemOpen
  \bibfield  {author} {\bibinfo {author} {\bibfnamefont {S.-Q.}\ \bibnamefont
  {Feng}}, \bibinfo {author} {\bibfnamefont {L.}~\bibnamefont {Pei}}, \bibinfo
  {author} {\bibfnamefont {F.}~\bibnamefont {Sun}}, \bibinfo {author}
  {\bibfnamefont {Y.}~\bibnamefont {Zhong}}, \ and\ \bibinfo {author}
  {\bibfnamefont {Z.-B.}\ \bibnamefont {Yin}},\ }\href {\doibase
  10.1088/1674-1137/42/5/054102} {\bibfield  {journal} {\bibinfo  {journal}
  {Chin. Phys. C}\ }\textbf {\bibinfo {volume} {42}},\ \bibinfo {pages}
  {054102} (\bibinfo {year} {2018})}\BibitemShut {NoStop}%
\bibitem [{\citenamefont {Skokov}\ \emph {et~al.}(2009)\citenamefont {Skokov},
  \citenamefont {Illarionov},\ and\ \citenamefont {Toneev}}]{Skokov:2009qp}%
  \BibitemOpen
  \bibfield  {author} {\bibinfo {author} {\bibfnamefont {V.}~\bibnamefont
  {Skokov}}, \bibinfo {author} {\bibfnamefont {A.~Y.}\ \bibnamefont
  {Illarionov}}, \ and\ \bibinfo {author} {\bibfnamefont {V.}~\bibnamefont
  {Toneev}},\ }\href {\doibase 10.1142/S0217751X09047570} {\bibfield  {journal}
  {\bibinfo  {journal} {Int. J. Mod. Phys. A}\ }\textbf {\bibinfo {volume}
  {24}},\ \bibinfo {pages} {5925} (\bibinfo {year} {2009})}\BibitemShut
  {NoStop}%
\bibitem [{\citenamefont {Tuchin}(2013)}]{Tuchin:2013ie}%
  \BibitemOpen
  \bibfield  {author} {\bibinfo {author} {\bibfnamefont {K.}~\bibnamefont
  {Tuchin}},\ }\href {\doibase 10.1155/2013/490495} {\bibfield  {journal}
  {\bibinfo  {journal} {Adv. High Energy Phys.}\ }\textbf {\bibinfo {volume}
  {2013}},\ \bibinfo {pages} {490495} (\bibinfo {year} {2013})}\BibitemShut
  {NoStop}%
\bibitem [{\citenamefont {Kharzeev}\ \emph {et~al.}(2016)\citenamefont
  {Kharzeev}, \citenamefont {Liao}, \citenamefont {Voloshin},\ and\
  \citenamefont {Wang}}]{Kharzeev:2015znc}%
  \BibitemOpen
  \bibfield  {author} {\bibinfo {author} {\bibfnamefont {D.~E.}\ \bibnamefont
  {Kharzeev}}, \bibinfo {author} {\bibfnamefont {J.}~\bibnamefont {Liao}},
  \bibinfo {author} {\bibfnamefont {S.~A.}\ \bibnamefont {Voloshin}}, \ and\
  \bibinfo {author} {\bibfnamefont {G.}~\bibnamefont {Wang}},\ }\href {\doibase
  10.1016/j.ppnp.2016.01.001} {\bibfield  {journal} {\bibinfo  {journal} {Prog.
  Part. Nucl. Phys.}\ }\textbf {\bibinfo {volume} {88}},\ \bibinfo {pages} {1}
  (\bibinfo {year} {2016})}\BibitemShut {NoStop}%
\bibitem [{\citenamefont {Guo}\ \emph {et~al.}(2019)\citenamefont {Guo},
  \citenamefont {Shi}, \citenamefont {Feng},\ and\ \citenamefont
  {Liao}}]{Guo:2019joy}%
  \BibitemOpen
  \bibfield  {author} {\bibinfo {author} {\bibfnamefont {Y.}~\bibnamefont
  {Guo}}, \bibinfo {author} {\bibfnamefont {S.}~\bibnamefont {Shi}}, \bibinfo
  {author} {\bibfnamefont {S.}~\bibnamefont {Feng}}, \ and\ \bibinfo {author}
  {\bibfnamefont {J.}~\bibnamefont {Liao}},\ }\href {\doibase
  10.1016/j.physletb.2019.134929} {\bibfield  {journal} {\bibinfo  {journal}
  {Phys. Lett. B}\ }\textbf {\bibinfo {volume} {798}},\ \bibinfo {pages}
  {134929} (\bibinfo {year} {2019})}\BibitemShut {NoStop}%
\bibitem [{\citenamefont {Zhao}\ and\ \citenamefont
  {Wang}(2019)}]{Zhao:2019hta}%
  \BibitemOpen
  \bibfield  {author} {\bibinfo {author} {\bibfnamefont {J.}~\bibnamefont
  {Zhao}}\ and\ \bibinfo {author} {\bibfnamefont {F.}~\bibnamefont {Wang}},\
  }\href {\doibase 10.1016/j.ppnp.2019.05.001} {\bibfield  {journal} {\bibinfo
  {journal} {Prog. Part. Nucl. Phys.}\ }\textbf {\bibinfo {volume} {107}},\
  \bibinfo {pages} {200} (\bibinfo {year} {2019})}\BibitemShut {NoStop}%
\bibitem [{\citenamefont {Chen}\ and\ \citenamefont
  {Feng}(2020)}]{Chen:2019qoe}%
  \BibitemOpen
  \bibfield  {author} {\bibinfo {author} {\bibfnamefont {B.-X.}\ \bibnamefont
  {Chen}}\ and\ \bibinfo {author} {\bibfnamefont {S.-Q.}\ \bibnamefont
  {Feng}},\ }\href {\doibase 10.1088/1674-1137/44/2/024104} {\bibfield
  {journal} {\bibinfo  {journal} {Chin. Phys. C}\ }\textbf {\bibinfo {volume}
  {44}},\ \bibinfo {pages} {024104} (\bibinfo {year} {2020})}\BibitemShut
  {NoStop}%
\bibitem [{\citenamefont {Tuchin}(2016)}]{Tuchin:2015oka}%
  \BibitemOpen
  \bibfield  {author} {\bibinfo {author} {\bibfnamefont {K.}~\bibnamefont
  {Tuchin}},\ }\href {\doibase 10.1103/PhysRevC.93.014905} {\bibfield
  {journal} {\bibinfo  {journal} {Phys. Rev. C}\ }\textbf {\bibinfo {volume}
  {93}},\ \bibinfo {pages} {014905} (\bibinfo {year} {2016})}\BibitemShut
  {NoStop}%
\bibitem [{\citenamefont {McLerran}\ and\ \citenamefont
  {Skokov}(2014)}]{McLerran:2013hla}%
  \BibitemOpen
  \bibfield  {author} {\bibinfo {author} {\bibfnamefont {L.}~\bibnamefont
  {McLerran}}\ and\ \bibinfo {author} {\bibfnamefont {V.}~\bibnamefont
  {Skokov}},\ }\href {\doibase 10.1016/j.nuclphysa.2014.05.008} {\bibfield
  {journal} {\bibinfo  {journal} {Nucl. Phys. A}\ }\textbf {\bibinfo {volume}
  {929}},\ \bibinfo {pages} {184} (\bibinfo {year} {2014})}\BibitemShut
  {NoStop}%
\bibitem [{\citenamefont {She}\ \emph {et~al.}(2018)\citenamefont {She},
  \citenamefont {Feng}, \citenamefont {Zhong},\ and\ \citenamefont
  {Yin}}]{She:2017icp}%
  \BibitemOpen
  \bibfield  {author} {\bibinfo {author} {\bibfnamefont {D.}~\bibnamefont
  {She}}, \bibinfo {author} {\bibfnamefont {S.-Q.}\ \bibnamefont {Feng}},
  \bibinfo {author} {\bibfnamefont {Y.}~\bibnamefont {Zhong}}, \ and\ \bibinfo
  {author} {\bibfnamefont {Z.-B.}\ \bibnamefont {Yin}},\ }\href {\doibase
  10.1140/epja/i2018-12481-x} {\bibfield  {journal} {\bibinfo  {journal} {Eur.
  Phys. J. A}\ }\textbf {\bibinfo {volume} {54}},\ \bibinfo {pages} {48}
  (\bibinfo {year} {2018})}\BibitemShut {NoStop}%
\bibitem [{\citenamefont {Sch\"afer}\ and\ \citenamefont
  {Teaney}(2009)}]{Schafer:2009dj}%
  \BibitemOpen
  \bibfield  {author} {\bibinfo {author} {\bibfnamefont {T.}~\bibnamefont
  {Sch\"afer}}\ and\ \bibinfo {author} {\bibfnamefont {D.}~\bibnamefont
  {Teaney}},\ }\href {\doibase 10.1088/0034-4885/72/12/126001} {\bibfield
  {journal} {\bibinfo  {journal} {Rept. Prog. Phys.}\ }\textbf {\bibinfo
  {volume} {72}},\ \bibinfo {pages} {126001} (\bibinfo {year}
  {2009})}\BibitemShut {NoStop}%
\bibitem [{\citenamefont {Andersen}\ \emph {et~al.}(2016)\citenamefont
  {Andersen}, \citenamefont {Naylor},\ and\ \citenamefont
  {Tranberg}}]{Andersen:2014xxa}%
  \BibitemOpen
  \bibfield  {author} {\bibinfo {author} {\bibfnamefont {J.~O.}\ \bibnamefont
  {Andersen}}, \bibinfo {author} {\bibfnamefont {W.~R.}\ \bibnamefont
  {Naylor}}, \ and\ \bibinfo {author} {\bibfnamefont {A.}~\bibnamefont
  {Tranberg}},\ }\href {\doibase 10.1103/RevModPhys.88.025001} {\bibfield
  {journal} {\bibinfo  {journal} {Rev. Mod. Phys.}\ }\textbf {\bibinfo {volume}
  {88}},\ \bibinfo {pages} {025001} (\bibinfo {year} {2016})}\BibitemShut
  {NoStop}%
\bibitem [{\citenamefont {Klevansky}\ and\ \citenamefont
  {Lemmer}(1989)}]{Klevansky:1989vi}%
  \BibitemOpen
  \bibfield  {author} {\bibinfo {author} {\bibfnamefont {S.~P.}\ \bibnamefont
  {Klevansky}}\ and\ \bibinfo {author} {\bibfnamefont {R.~H.}\ \bibnamefont
  {Lemmer}},\ }\href {\doibase 10.1103/PhysRevD.39.3478} {\bibfield  {journal}
  {\bibinfo  {journal} {Phys. Rev. D}\ }\textbf {\bibinfo {volume} {39}},\
  \bibinfo {pages} {3478} (\bibinfo {year} {1989})}\BibitemShut {NoStop}%
\bibitem [{\citenamefont {Gatto}\ and\ \citenamefont
  {Ruggieri}(2011)}]{Gatto:2010pt}%
  \BibitemOpen
  \bibfield  {author} {\bibinfo {author} {\bibfnamefont {R.}~\bibnamefont
  {Gatto}}\ and\ \bibinfo {author} {\bibfnamefont {M.}~\bibnamefont
  {Ruggieri}},\ }\href {\doibase 10.1103/PhysRevD.83.034016} {\bibfield
  {journal} {\bibinfo  {journal} {Phys. Rev. D}\ }\textbf {\bibinfo {volume}
  {83}},\ \bibinfo {pages} {034016} (\bibinfo {year} {2011})}\BibitemShut
  {NoStop}%
\bibitem [{\citenamefont {Miransky}\ and\ \citenamefont
  {Shovkovy}(2015)}]{Miransky:2015ava}%
  \BibitemOpen
  \bibfield  {author} {\bibinfo {author} {\bibfnamefont {V.~A.}\ \bibnamefont
  {Miransky}}\ and\ \bibinfo {author} {\bibfnamefont {I.~A.}\ \bibnamefont
  {Shovkovy}},\ }\href {\doibase 10.1016/j.physrep.2015.02.003} {\bibfield
  {journal} {\bibinfo  {journal} {Phys. Rept.}\ }\textbf {\bibinfo {volume}
  {576}},\ \bibinfo {pages} {1} (\bibinfo {year} {2015})}\BibitemShut {NoStop}%
\bibitem [{\citenamefont {Boomsma}\ and\ \citenamefont
  {Boer}(2010)}]{Boomsma:2009yk}%
  \BibitemOpen
  \bibfield  {author} {\bibinfo {author} {\bibfnamefont {J.~K.}\ \bibnamefont
  {Boomsma}}\ and\ \bibinfo {author} {\bibfnamefont {D.}~\bibnamefont {Boer}},\
  }\href {\doibase 10.1103/PhysRevD.81.074005} {\bibfield  {journal} {\bibinfo
  {journal} {Phys. Rev. D}\ }\textbf {\bibinfo {volume} {81}},\ \bibinfo
  {pages} {074005} (\bibinfo {year} {2010})}\BibitemShut {NoStop}%
\bibitem [{\citenamefont {Gatto}\ and\ \citenamefont
  {Ruggieri}(2013)}]{Gatto:2012sp}%
  \BibitemOpen
  \bibfield  {author} {\bibinfo {author} {\bibfnamefont {R.}~\bibnamefont
  {Gatto}}\ and\ \bibinfo {author} {\bibfnamefont {M.}~\bibnamefont
  {Ruggieri}},\ }\href {\doibase 10.1007/978-3-642-37305-3_4} {\bibfield
  {journal} {\bibinfo  {journal} {Lect. Notes Phys.}\ }\textbf {\bibinfo
  {volume} {871}},\ \bibinfo {pages} {87} (\bibinfo {year} {2013})}\BibitemShut
  {NoStop}%
\bibitem [{\citenamefont {Chatterjee}\ \emph {et~al.}(2015)\citenamefont
  {Chatterjee}, \citenamefont {Mishra},\ and\ \citenamefont
  {Mishra}}]{Chatterjee:2014csa}%
  \BibitemOpen
  \bibfield  {author} {\bibinfo {author} {\bibfnamefont {B.}~\bibnamefont
  {Chatterjee}}, \bibinfo {author} {\bibfnamefont {H.}~\bibnamefont {Mishra}},
  \ and\ \bibinfo {author} {\bibfnamefont {A.}~\bibnamefont {Mishra}},\ }\href
  {\doibase 10.1103/PhysRevD.91.034031} {\bibfield  {journal} {\bibinfo
  {journal} {Phys. Rev. D}\ }\textbf {\bibinfo {volume} {91}},\ \bibinfo
  {pages} {034031} (\bibinfo {year} {2015})}\BibitemShut {NoStop}%
\bibitem [{\citenamefont {Chatrchyan}\ \emph {et~al.}(2012)\citenamefont
  {Chatrchyan} \emph {et~al.}}]{CMS:2012sap}%
  \BibitemOpen
  \bibfield  {author} {\bibinfo {author} {\bibfnamefont {S.}~\bibnamefont
  {Chatrchyan}} \emph {et~al.} (\bibinfo {collaboration} {CMS}),\ }\href
  {\doibase 10.1007/JHEP05(2012)055} {\bibfield  {journal} {\bibinfo  {journal}
  {JHEP}\ }\textbf {\bibinfo {volume} {05}},\ \bibinfo {pages} {055} (\bibinfo
  {year} {2012})}\BibitemShut {NoStop}%
\bibitem [{\citenamefont {Bali}\ \emph
  {et~al.}(2012{\natexlab{a}})\citenamefont {Bali}, \citenamefont {Bruckmann},
  \citenamefont {Endrodi}, \citenamefont {Fodor}, \citenamefont {Katz},\ and\
  \citenamefont {Schafer}}]{Bali:2012zg}%
  \BibitemOpen
  \bibfield  {author} {\bibinfo {author} {\bibfnamefont {G.~S.}\ \bibnamefont
  {Bali}}, \bibinfo {author} {\bibfnamefont {F.}~\bibnamefont {Bruckmann}},
  \bibinfo {author} {\bibfnamefont {G.}~\bibnamefont {Endrodi}}, \bibinfo
  {author} {\bibfnamefont {Z.}~\bibnamefont {Fodor}}, \bibinfo {author}
  {\bibfnamefont {S.~D.}\ \bibnamefont {Katz}}, \ and\ \bibinfo {author}
  {\bibfnamefont {A.}~\bibnamefont {Schafer}},\ }\href {\doibase
  10.1103/PhysRevD.86.071502} {\bibfield  {journal} {\bibinfo  {journal} {Phys.
  Rev. D}\ }\textbf {\bibinfo {volume} {86}},\ \bibinfo {pages} {071502}
  (\bibinfo {year} {2012}{\natexlab{a}})}\BibitemShut {NoStop}%
\bibitem [{\citenamefont {Bruckmann}\ \emph {et~al.}(2013)\citenamefont
  {Bruckmann}, \citenamefont {Endrodi},\ and\ \citenamefont
  {Kovacs}}]{Bruckmann:2013oba}%
  \BibitemOpen
  \bibfield  {author} {\bibinfo {author} {\bibfnamefont {F.}~\bibnamefont
  {Bruckmann}}, \bibinfo {author} {\bibfnamefont {G.}~\bibnamefont {Endrodi}},
  \ and\ \bibinfo {author} {\bibfnamefont {T.~G.}\ \bibnamefont {Kovacs}},\
  }\href {\doibase 10.1007/JHEP04(2013)112} {\bibfield  {journal} {\bibinfo
  {journal} {JHEP}\ }\textbf {\bibinfo {volume} {04}},\ \bibinfo {pages} {112}
  (\bibinfo {year} {2013})}\BibitemShut {NoStop}%
\bibitem [{\citenamefont {Endr\"odi}(2013)}]{Endrodi:2013cs}%
  \BibitemOpen
  \bibfield  {author} {\bibinfo {author} {\bibfnamefont {G.}~\bibnamefont
  {Endr\"odi}},\ }\href {\doibase 10.1007/JHEP04(2013)023} {\bibfield
  {journal} {\bibinfo  {journal} {JHEP}\ }\textbf {\bibinfo {volume} {04}},\
  \bibinfo {pages} {023} (\bibinfo {year} {2013})}\BibitemShut {NoStop}%
\bibitem [{\citenamefont {Sasaki}\ and\ \citenamefont
  {Redlich}(2010)}]{Sasaki:2008um}%
  \BibitemOpen
  \bibfield  {author} {\bibinfo {author} {\bibfnamefont {C.}~\bibnamefont
  {Sasaki}}\ and\ \bibinfo {author} {\bibfnamefont {K.}~\bibnamefont
  {Redlich}},\ }\href {\doibase 10.1016/j.nuclphysa.2009.11.005} {\bibfield
  {journal} {\bibinfo  {journal} {Nucl. Phys. A}\ }\textbf {\bibinfo {volume}
  {832}},\ \bibinfo {pages} {62} (\bibinfo {year} {2010})}\BibitemShut
  {NoStop}%
\bibitem [{\citenamefont {Soloveva}\ \emph {et~al.}(2022)\citenamefont
  {Soloveva}, \citenamefont {Aichelin},\ and\ \citenamefont
  {Bratkovskaya}}]{Soloveva:2021quj}%
  \BibitemOpen
  \bibfield  {author} {\bibinfo {author} {\bibfnamefont {O.}~\bibnamefont
  {Soloveva}}, \bibinfo {author} {\bibfnamefont {J.}~\bibnamefont {Aichelin}},
  \ and\ \bibinfo {author} {\bibfnamefont {E.}~\bibnamefont {Bratkovskaya}},\
  }\href {\doibase 10.1103/PhysRevD.105.054011} {\bibfield  {journal} {\bibinfo
   {journal} {Phys. Rev. D}\ }\textbf {\bibinfo {volume} {105}},\ \bibinfo
  {pages} {054011} (\bibinfo {year} {2022})}\BibitemShut {NoStop}%
\bibitem [{\citenamefont {Mykhaylova}\ and\ \citenamefont
  {Sasaki}(2021)}]{Mykhaylova:2020pfk}%
  \BibitemOpen
  \bibfield  {author} {\bibinfo {author} {\bibfnamefont {V.}~\bibnamefont
  {Mykhaylova}}\ and\ \bibinfo {author} {\bibfnamefont {C.}~\bibnamefont
  {Sasaki}},\ }\href {\doibase 10.1103/PhysRevD.103.014007} {\bibfield
  {journal} {\bibinfo  {journal} {Phys. Rev. D}\ }\textbf {\bibinfo {volume}
  {103}},\ \bibinfo {pages} {014007} (\bibinfo {year} {2021})}\BibitemShut
  {NoStop}%
\bibitem [{\citenamefont {Hatsuda}\ and\ \citenamefont
  {Kunihiro}(1994)}]{Hatsuda:1994pi}%
  \BibitemOpen
  \bibfield  {author} {\bibinfo {author} {\bibfnamefont {T.}~\bibnamefont
  {Hatsuda}}\ and\ \bibinfo {author} {\bibfnamefont {T.}~\bibnamefont
  {Kunihiro}},\ }\href {\doibase 10.1016/0370-1573(94)90022-1} {\bibfield
  {journal} {\bibinfo  {journal} {Phys. Rept.}\ }\textbf {\bibinfo {volume}
  {247}},\ \bibinfo {pages} {221} (\bibinfo {year} {1994})}\BibitemShut
  {NoStop}%
\bibitem [{\citenamefont {Buballa}(2005)}]{Buballa:2003qv}%
  \BibitemOpen
  \bibfield  {author} {\bibinfo {author} {\bibfnamefont {M.}~\bibnamefont
  {Buballa}},\ }\href {\doibase 10.1016/j.physrep.2004.11.004} {\bibfield
  {journal} {\bibinfo  {journal} {Phys. Rept.}\ }\textbf {\bibinfo {volume}
  {407}},\ \bibinfo {pages} {205} (\bibinfo {year} {2005})}\BibitemShut
  {NoStop}%
\bibitem [{\citenamefont {Li}\ and\ \citenamefont {Yee}(2018)}]{Li:2017tgi}%
  \BibitemOpen
  \bibfield  {author} {\bibinfo {author} {\bibfnamefont {S.}~\bibnamefont
  {Li}}\ and\ \bibinfo {author} {\bibfnamefont {H.-U.}\ \bibnamefont {Yee}},\
  }\href {\doibase 10.1103/PhysRevD.97.056024} {\bibfield  {journal} {\bibinfo
  {journal} {Phys. Rev. D}\ }\textbf {\bibinfo {volume} {97}},\ \bibinfo
  {pages} {056024} (\bibinfo {year} {2018})}\BibitemShut {NoStop}%
\bibitem [{\citenamefont {Nam}\ and\ \citenamefont {Kao}(2013)}]{Nam:2013fpa}%
  \BibitemOpen
  \bibfield  {author} {\bibinfo {author} {\bibfnamefont {S.-i.}\ \bibnamefont
  {Nam}}\ and\ \bibinfo {author} {\bibfnamefont {C.-W.}\ \bibnamefont {Kao}},\
  }\href {\doibase 10.1103/PhysRevD.87.114003} {\bibfield  {journal} {\bibinfo
  {journal} {Phys. Rev. D}\ }\textbf {\bibinfo {volume} {87}},\ \bibinfo
  {pages} {114003} (\bibinfo {year} {2013})}\BibitemShut {NoStop}%
\bibitem [{\citenamefont {Alford}\ \emph {et~al.}(2014)\citenamefont {Alford},
  \citenamefont {Nishimura},\ and\ \citenamefont {Sedrakian}}]{Alford:2014doa}%
  \BibitemOpen
  \bibfield  {author} {\bibinfo {author} {\bibfnamefont {M.~G.}\ \bibnamefont
  {Alford}}, \bibinfo {author} {\bibfnamefont {H.}~\bibnamefont {Nishimura}}, \
  and\ \bibinfo {author} {\bibfnamefont {A.}~\bibnamefont {Sedrakian}},\ }\href
  {\doibase 10.1103/PhysRevC.90.055205} {\bibfield  {journal} {\bibinfo
  {journal} {Phys. Rev. C}\ }\textbf {\bibinfo {volume} {90}},\ \bibinfo
  {pages} {055205} (\bibinfo {year} {2014})}\BibitemShut {NoStop}%
\bibitem [{\citenamefont {Tuchin}(2012)}]{Tuchin:2011jw}%
  \BibitemOpen
  \bibfield  {author} {\bibinfo {author} {\bibfnamefont {K.}~\bibnamefont
  {Tuchin}},\ }\href {\doibase 10.1088/0954-3899/39/2/025010} {\bibfield
  {journal} {\bibinfo  {journal} {J. Phys. G}\ }\textbf {\bibinfo {volume}
  {39}},\ \bibinfo {pages} {025010} (\bibinfo {year} {2012})}\BibitemShut
  {NoStop}%
\bibitem [{\citenamefont {Hattori}\ \emph {et~al.}(2017)\citenamefont
  {Hattori}, \citenamefont {Huang}, \citenamefont {Rischke},\ and\
  \citenamefont {Satow}}]{Hattori:2017qih}%
  \BibitemOpen
  \bibfield  {author} {\bibinfo {author} {\bibfnamefont {K.}~\bibnamefont
  {Hattori}}, \bibinfo {author} {\bibfnamefont {X.-G.}\ \bibnamefont {Huang}},
  \bibinfo {author} {\bibfnamefont {D.~H.}\ \bibnamefont {Rischke}}, \ and\
  \bibinfo {author} {\bibfnamefont {D.}~\bibnamefont {Satow}},\ }\href
  {\doibase 10.1103/PhysRevD.96.094009} {\bibfield  {journal} {\bibinfo
  {journal} {Phys. Rev. D}\ }\textbf {\bibinfo {volume} {96}},\ \bibinfo
  {pages} {094009} (\bibinfo {year} {2017})}\BibitemShut {NoStop}%
\bibitem [{\citenamefont {Huang}\ \emph {et~al.}(2010)\citenamefont {Huang},
  \citenamefont {Huang}, \citenamefont {Rischke},\ and\ \citenamefont
  {Sedrakian}}]{Huang:2009ue}%
  \BibitemOpen
  \bibfield  {author} {\bibinfo {author} {\bibfnamefont {X.-G.}\ \bibnamefont
  {Huang}}, \bibinfo {author} {\bibfnamefont {M.}~\bibnamefont {Huang}},
  \bibinfo {author} {\bibfnamefont {D.~H.}\ \bibnamefont {Rischke}}, \ and\
  \bibinfo {author} {\bibfnamefont {A.}~\bibnamefont {Sedrakian}},\ }\href
  {\doibase 10.1103/PhysRevD.81.045015} {\bibfield  {journal} {\bibinfo
  {journal} {Phys. Rev. D}\ }\textbf {\bibinfo {volume} {81}},\ \bibinfo
  {pages} {045015} (\bibinfo {year} {2010})}\BibitemShut {NoStop}%
\bibitem [{\citenamefont {Huang}\ \emph {et~al.}(2011)\citenamefont {Huang},
  \citenamefont {Sedrakian},\ and\ \citenamefont {Rischke}}]{Huang:2011dc}%
  \BibitemOpen
  \bibfield  {author} {\bibinfo {author} {\bibfnamefont {X.-G.}\ \bibnamefont
  {Huang}}, \bibinfo {author} {\bibfnamefont {A.}~\bibnamefont {Sedrakian}}, \
  and\ \bibinfo {author} {\bibfnamefont {D.~H.}\ \bibnamefont {Rischke}},\
  }\href {\doibase 10.1016/j.aop.2011.08.001} {\bibfield  {journal} {\bibinfo
  {journal} {Annals Phys.}\ }\textbf {\bibinfo {volume} {326}},\ \bibinfo
  {pages} {3075} (\bibinfo {year} {2011})}\BibitemShut {NoStop}%
\bibitem [{\citenamefont {Mohanty}\ \emph {et~al.}(2019)\citenamefont
  {Mohanty}, \citenamefont {Dash},\ and\ \citenamefont
  {Roy}}]{Mohanty:2018eja}%
  \BibitemOpen
  \bibfield  {author} {\bibinfo {author} {\bibfnamefont {P.}~\bibnamefont
  {Mohanty}}, \bibinfo {author} {\bibfnamefont {A.}~\bibnamefont {Dash}}, \
  and\ \bibinfo {author} {\bibfnamefont {V.}~\bibnamefont {Roy}},\ }\href
  {\doibase 10.1140/epja/i2019-12705-7} {\bibfield  {journal} {\bibinfo
  {journal} {Eur. Phys. J. A}\ }\textbf {\bibinfo {volume} {55}},\ \bibinfo
  {pages} {35} (\bibinfo {year} {2019})}\BibitemShut {NoStop}%
\bibitem [{\citenamefont {Bali}\ \emph
  {et~al.}(2012{\natexlab{b}})\citenamefont {Bali}, \citenamefont {Bruckmann},
  \citenamefont {Endrodi}, \citenamefont {Fodor}, \citenamefont {Katz},
  \citenamefont {Krieg}, \citenamefont {Schafer},\ and\ \citenamefont
  {Szabo}}]{Bali:2011qj}%
  \BibitemOpen
  \bibfield  {author} {\bibinfo {author} {\bibfnamefont {G.~S.}\ \bibnamefont
  {Bali}}, \bibinfo {author} {\bibfnamefont {F.}~\bibnamefont {Bruckmann}},
  \bibinfo {author} {\bibfnamefont {G.}~\bibnamefont {Endrodi}}, \bibinfo
  {author} {\bibfnamefont {Z.}~\bibnamefont {Fodor}}, \bibinfo {author}
  {\bibfnamefont {S.~D.}\ \bibnamefont {Katz}}, \bibinfo {author}
  {\bibfnamefont {S.}~\bibnamefont {Krieg}}, \bibinfo {author} {\bibfnamefont
  {A.}~\bibnamefont {Schafer}}, \ and\ \bibinfo {author} {\bibfnamefont
  {K.~K.}\ \bibnamefont {Szabo}},\ }\href {\doibase 10.1007/JHEP02(2012)044}
  {\bibfield  {journal} {\bibinfo  {journal} {JHEP}\ }\textbf {\bibinfo
  {volume} {02}},\ \bibinfo {pages} {044} (\bibinfo {year}
  {2012}{\natexlab{b}})}\BibitemShut {NoStop}%
\bibitem [{\citenamefont {Nambu}\ and\ \citenamefont
  {Jona-Lasinio}(1961)}]{Nambu:1961tp}%
  \BibitemOpen
  \bibfield  {author} {\bibinfo {author} {\bibfnamefont {Y.}~\bibnamefont
  {Nambu}}\ and\ \bibinfo {author} {\bibfnamefont {G.}~\bibnamefont
  {Jona-Lasinio}},\ }\href {\doibase 10.1103/PhysRev.122.345} {\bibfield
  {journal} {\bibinfo  {journal} {Phys. Rev.}\ }\textbf {\bibinfo {volume}
  {122}},\ \bibinfo {pages} {345} (\bibinfo {year} {1961})}\BibitemShut
  {NoStop}%
\bibitem [{\citenamefont {Menezes}\ \emph {et~al.}(2009)\citenamefont
  {Menezes}, \citenamefont {Benghi~Pinto}, \citenamefont {Avancini},
  \citenamefont {Perez~Martinez},\ and\ \citenamefont
  {Providencia}}]{Menezes:2008qt}%
  \BibitemOpen
  \bibfield  {author} {\bibinfo {author} {\bibfnamefont {D.~P.}\ \bibnamefont
  {Menezes}}, \bibinfo {author} {\bibfnamefont {M.}~\bibnamefont
  {Benghi~Pinto}}, \bibinfo {author} {\bibfnamefont {S.~S.}\ \bibnamefont
  {Avancini}}, \bibinfo {author} {\bibfnamefont {A.}~\bibnamefont
  {Perez~Martinez}}, \ and\ \bibinfo {author} {\bibfnamefont {C.}~\bibnamefont
  {Providencia}},\ }\href {\doibase 10.1103/PhysRevC.79.035807} {\bibfield
  {journal} {\bibinfo  {journal} {Phys. Rev. C}\ }\textbf {\bibinfo {volume}
  {79}},\ \bibinfo {pages} {035807} (\bibinfo {year} {2009})}\BibitemShut
  {NoStop}%
\bibitem [{\citenamefont {Ferrari}\ \emph {et~al.}(2012)\citenamefont
  {Ferrari}, \citenamefont {Garcia},\ and\ \citenamefont
  {Pinto}}]{Ferrari:2012yw}%
  \BibitemOpen
  \bibfield  {author} {\bibinfo {author} {\bibfnamefont {G.~N.}\ \bibnamefont
  {Ferrari}}, \bibinfo {author} {\bibfnamefont {A.~F.}\ \bibnamefont {Garcia}},
  \ and\ \bibinfo {author} {\bibfnamefont {M.~B.}\ \bibnamefont {Pinto}},\
  }\href {\doibase 10.1103/PhysRevD.86.096005} {\bibfield  {journal} {\bibinfo
  {journal} {Phys. Rev. D}\ }\textbf {\bibinfo {volume} {86}},\ \bibinfo
  {pages} {096005} (\bibinfo {year} {2012})}\BibitemShut {NoStop}%
\bibitem [{\citenamefont {Farias}\ \emph {et~al.}(2014)\citenamefont {Farias},
  \citenamefont {Gomes}, \citenamefont {Krein},\ and\ \citenamefont
  {Pinto}}]{Farias:2014eca}%
  \BibitemOpen
  \bibfield  {author} {\bibinfo {author} {\bibfnamefont {R.~L.~S.}\
  \bibnamefont {Farias}}, \bibinfo {author} {\bibfnamefont {K.~P.}\
  \bibnamefont {Gomes}}, \bibinfo {author} {\bibfnamefont {G.~I.}\ \bibnamefont
  {Krein}}, \ and\ \bibinfo {author} {\bibfnamefont {M.~B.}\ \bibnamefont
  {Pinto}},\ }\href {\doibase 10.1103/PhysRevC.90.025203} {\bibfield  {journal}
  {\bibinfo  {journal} {Phys. Rev. C}\ }\textbf {\bibinfo {volume} {90}},\
  \bibinfo {pages} {025203} (\bibinfo {year} {2014})}\BibitemShut {NoStop}%
\bibitem [{\citenamefont {Farias}\ \emph {et~al.}(2017)\citenamefont {Farias},
  \citenamefont {Timoteo}, \citenamefont {Avancini}, \citenamefont {Pinto},\
  and\ \citenamefont {Krein}}]{Farias:2016gmy}%
  \BibitemOpen
  \bibfield  {author} {\bibinfo {author} {\bibfnamefont {R.~L.~S.}\
  \bibnamefont {Farias}}, \bibinfo {author} {\bibfnamefont {V.~S.}\
  \bibnamefont {Timoteo}}, \bibinfo {author} {\bibfnamefont {S.~S.}\
  \bibnamefont {Avancini}}, \bibinfo {author} {\bibfnamefont {M.~B.}\
  \bibnamefont {Pinto}}, \ and\ \bibinfo {author} {\bibfnamefont
  {G.}~\bibnamefont {Krein}},\ }\href {\doibase 10.1140/epja/i2017-12320-8}
  {\bibfield  {journal} {\bibinfo  {journal} {Eur. Phys. J. A}\ }\textbf
  {\bibinfo {volume} {53}},\ \bibinfo {pages} {101} (\bibinfo {year}
  {2017})}\BibitemShut {NoStop}%
\bibitem [{\citenamefont {Ayala}\ \emph {et~al.}(2014)\citenamefont {Ayala},
  \citenamefont {Loewe}, \citenamefont {Mizher},\ and\ \citenamefont
  {Zamora}}]{Ayala:2014iba}%
  \BibitemOpen
  \bibfield  {author} {\bibinfo {author} {\bibfnamefont {A.}~\bibnamefont
  {Ayala}}, \bibinfo {author} {\bibfnamefont {M.}~\bibnamefont {Loewe}},
  \bibinfo {author} {\bibfnamefont {A.~J.}\ \bibnamefont {Mizher}}, \ and\
  \bibinfo {author} {\bibfnamefont {R.}~\bibnamefont {Zamora}},\ }\href
  {\doibase 10.1103/PhysRevD.90.036001} {\bibfield  {journal} {\bibinfo
  {journal} {Phys. Rev. D}\ }\textbf {\bibinfo {volume} {90}},\ \bibinfo
  {pages} {036001} (\bibinfo {year} {2014})}\BibitemShut {NoStop}%
\bibitem [{\citenamefont {Ayala}\ \emph {et~al.}(2015)\citenamefont {Ayala},
  \citenamefont {Loewe},\ and\ \citenamefont {Zamora}}]{Ayala:2014gwa}%
  \BibitemOpen
  \bibfield  {author} {\bibinfo {author} {\bibfnamefont {A.}~\bibnamefont
  {Ayala}}, \bibinfo {author} {\bibfnamefont {M.}~\bibnamefont {Loewe}}, \ and\
  \bibinfo {author} {\bibfnamefont {R.}~\bibnamefont {Zamora}},\ }\href
  {\doibase 10.1103/PhysRevD.91.016002} {\bibfield  {journal} {\bibinfo
  {journal} {Phys. Rev. D}\ }\textbf {\bibinfo {volume} {91}},\ \bibinfo
  {pages} {016002} (\bibinfo {year} {2015})}\BibitemShut {NoStop}%
\bibitem [{\citenamefont {Miransky}\ and\ \citenamefont
  {Shovkovy}(2002)}]{Miransky:2002rp}%
  \BibitemOpen
  \bibfield  {author} {\bibinfo {author} {\bibfnamefont {V.~A.}\ \bibnamefont
  {Miransky}}\ and\ \bibinfo {author} {\bibfnamefont {I.~A.}\ \bibnamefont
  {Shovkovy}},\ }\href {\doibase 10.1103/PhysRevD.66.045006} {\bibfield
  {journal} {\bibinfo  {journal} {Phys. Rev. D}\ }\textbf {\bibinfo {volume}
  {66}},\ \bibinfo {pages} {045006} (\bibinfo {year} {2002})}\BibitemShut
  {NoStop}%
\bibitem [{\citenamefont {Su}\ and\ \citenamefont {Wen}(2021)}]{Su:2021cda}%
  \BibitemOpen
  \bibfield  {author} {\bibinfo {author} {\bibfnamefont {S.-Z.}\ \bibnamefont
  {Su}}\ and\ \bibinfo {author} {\bibfnamefont {X.-J.}\ \bibnamefont {Wen}},\
  }\href {\doibase 10.1088/1361-6471/abfbc3} {\bibfield  {journal} {\bibinfo
  {journal} {J. Phys. G}\ }\textbf {\bibinfo {volume} {48}},\ \bibinfo {pages}
  {075004} (\bibinfo {year} {2021})}\BibitemShut {NoStop}%
\bibitem [{\citenamefont {Fodor}\ and\ \citenamefont
  {Katz}(2004)}]{Fodor:2004nz}%
  \BibitemOpen
  \bibfield  {author} {\bibinfo {author} {\bibfnamefont {Z.}~\bibnamefont
  {Fodor}}\ and\ \bibinfo {author} {\bibfnamefont {S.~D.}\ \bibnamefont
  {Katz}},\ }\href {\doibase 10.1088/1126-6708/2004/04/050} {\bibfield
  {journal} {\bibinfo  {journal} {JHEP}\ }\textbf {\bibinfo {volume} {04}},\
  \bibinfo {pages} {050} (\bibinfo {year} {2004})}\BibitemShut {NoStop}%
\bibitem [{\citenamefont {Costa}\ \emph {et~al.}(2015)\citenamefont {Costa},
  \citenamefont {Ferreira}, \citenamefont {Menezes}, \citenamefont {Moreira},\
  and\ \citenamefont {Provid\^encia}}]{Costa:2015bza}%
  \BibitemOpen
  \bibfield  {author} {\bibinfo {author} {\bibfnamefont {P.}~\bibnamefont
  {Costa}}, \bibinfo {author} {\bibfnamefont {M.}~\bibnamefont {Ferreira}},
  \bibinfo {author} {\bibfnamefont {D.~P.}\ \bibnamefont {Menezes}}, \bibinfo
  {author} {\bibfnamefont {J.~a.}\ \bibnamefont {Moreira}}, \ and\ \bibinfo
  {author} {\bibfnamefont {C.}~\bibnamefont {Provid\^encia}},\ }\href {\doibase
  10.1103/PhysRevD.92.036012} {\bibfield  {journal} {\bibinfo  {journal} {Phys.
  Rev. D}\ }\textbf {\bibinfo {volume} {92}},\ \bibinfo {pages} {036012}
  (\bibinfo {year} {2015})}\BibitemShut {NoStop}%
\bibitem [{\citenamefont {Ghosh}\ \emph {et~al.}(2019)\citenamefont {Ghosh},
  \citenamefont {Chatterjee}, \citenamefont {Mohanty}, \citenamefont
  {Mukharjee},\ and\ \citenamefont {Mishra}}]{Ghosh:2018cxb}%
  \BibitemOpen
  \bibfield  {author} {\bibinfo {author} {\bibfnamefont {S.}~\bibnamefont
  {Ghosh}}, \bibinfo {author} {\bibfnamefont {B.}~\bibnamefont {Chatterjee}},
  \bibinfo {author} {\bibfnamefont {P.}~\bibnamefont {Mohanty}}, \bibinfo
  {author} {\bibfnamefont {A.}~\bibnamefont {Mukharjee}}, \ and\ \bibinfo
  {author} {\bibfnamefont {H.}~\bibnamefont {Mishra}},\ }\href {\doibase
  10.1103/PhysRevD.100.034024} {\bibfield  {journal} {\bibinfo  {journal}
  {Phys. Rev. D}\ }\textbf {\bibinfo {volume} {100}},\ \bibinfo {pages}
  {034024} (\bibinfo {year} {2019})}\BibitemShut {NoStop}%
\end{thebibliography}%

\end{document}